\documentclass[aps,prb,twocolumn,showpacs]{revtex4-1}

\usepackage{graphicx}
\usepackage{dcolumn}
\usepackage{bm}
\usepackage{amsmath}
\usepackage{amssymb}
\usepackage{hyperref}
 \usepackage{epstopdf}

\begin{document}
\title{Occupation probabilities and current densities of bulk and edge states of a Floquet topological insulator}
\author{Hossein Dehghani}
\author{Aditi Mitra}
\affiliation{Department of Physics, New York University, 4 Washington Place, New York, NY 10003, USA
}
\date{\today}

\begin{abstract}
Results are presented for the occupation probabilities and current densities of bulk and edge states of half-filled
graphene in a cylindrical geometry, and irradiated by a circularly polarized laser. It is assumed that the system is closed, and that the laser
has been switched on as a quench. 
Laser parameters corresponding to some representative
topological phases are studied: 
one where the Chern number of the Floquet bands equals the number of chiral edge modes, 
a second where anomalous edge states
appear in the Floquet Brillouin zone boundaries, and a third where the 
Chern number is zero, yet topological edge states appear at the center and
boundaries of the Floquet Brillouin zone. Qualitative differences are found for 
the high frequency off-resonant and low frequency on-resonant laser with
edge states arising due to resonant processes occupied
with a high effective temperature on the one hand, while edge states arising due to off-resonant 
processes occupied with a low effective temperature on the other.
For an ideal half-filled system where only one of the bands in the Floquet
Brillouin zone is occupied and the other empty, particle-hole and inversion symmetry of the Floquet Hamiltonian 
implies zero current density. However the laser switch-on protocol breaks the inversion symmetry, 
resulting in a net cylindrical sheet of current 
density at steady-state. Due to the underlying chirality of the system, this current density profile is associated with a net 
charge imbalance between the top and bottom of the cylinders.
\end{abstract}

\pacs{73.43.-f, 03.65.Vf, 72.80.Vp}
\maketitle

\section{Introduction}

Topological systems are characterized by edge excitations that are remarkably robust
to perturbations. They arise due to a bulk-boundary correspondence, where
geometric properties of the bulk band-structure 
control the nature of 
excitations at the edge when the system is placed in a confined geometry.
Thus perturbations that cannot affect the bulk topological properties, cannot perturb the 
edge states either. For an integer quantum Hall system for example, bulk bands have a non-zero
Chern number $C$, which also equals the number
of chiral edge modes~\cite{TKNN,Bellissard94,Avron94}. The topological nature of the system
is responsible for the highly precise quantization of the Hall conductance at $C e^2/h$.~\cite{Klitzing80}

Chern insulators are topological insulators (TIs) which show quantum Hall physics in
the absence of a magnetic field, where time-reversal symmetry is broken by introducing complex
hopping amplitudes~\cite{Haldane88}. This can be achieved by doping with magnetic impurities~\cite{Hasan12}. 
Chern insulators can also be realized by the application of a circularly polarized
laser~\cite{Oka09,Inoue10,Kitagawa10,Lindner11}, where TIs arising out of such time-periodic
perturbations are referred to as Floquet TIs (FTIs)~\cite{Lindner11}.

The field of FTIs has grown in recent years because of
several experimental realizations ranging from periodically shaken lattices of cold-atomic gases~\cite{Esslinger14}, to
graphene~\cite{Karch10,Karch11}, Dirac fermions on the surface of 3D TIs~\cite{Gedik13} under external irradiation,
and photonic systems~\cite{Segev13,Hafezi13}. In fact FTIs are extremely rich, showing a variety of topological phases
as the amplitude, frequency, and polarization of the periodic drive is varied~\cite{Rudner13,Kundu14,Carpentier14,Dehghani15a}.
\begin{figure}
\includegraphics[height=9cm,width=9cm,keepaspectratio]{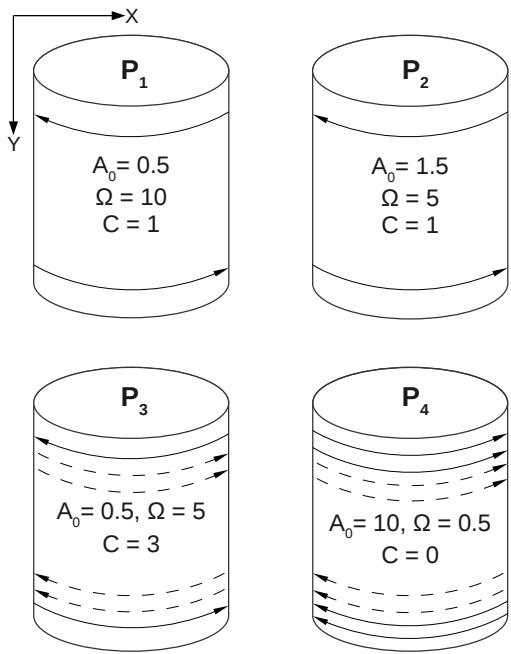}\\
\caption{
Sketch of the four topological phases studied, and labeled by $P_{1,2,3,4}$. 
Laser frequency $\Omega$ is in units of the hopping strength $t_h$ with $6t_h$ being the
bandwidth of graphene. Laser amplitude $A_0$ is in units of the lattice spacing.
Solid lines are edge states from the center of the FBZ, while dashed lines are
edge states from the boundary of the FBZ. The arrows indicate the chirality of the edge states.
}
\label{fig1}
\end{figure}

The topological properties of time-periodic Hamiltonians are extracted by studying the
spectral properties of an effective time-independent Hamiltonian known as the
Floquet Hamiltonian,~\cite{Sambe73} which captures the time-evolution of the system over
one period. Since energy is not conserved upto integer multiples of the driving frequency, the eigen-energies
of the Floquet Hamiltonian are known as quasi-energies. Much like spatially periodic systems,
here too, the Floquet description leads to an over-counting where 
quasi-energies separated by integer multiples of the laser frequency represent identical
eigen-modes. Thus in order to avoid this over-counting, one restricts the quasi-spectrum to one frequency of
the periodic drive, the so called Floquet Brillouin zone (FBZ). 

Analysis of the Floquet quasi-energies and quasi-modes shows that not only can
FTIs be used to realize conventional Chern insulators,~\cite{Oka09,Esslinger14} they also have 
unique properties coming from the fact that the energy is not conserved by integer multiples of the periodic drive.
As a result of this, the Chern number of the Floquet bands simply inform us of the difference between the  number
of chiral edge modes above and below the Floquet band. Effective (2+1) D topological invariants need to
be defined to account for the the periodicity in the additional temporal direction, and to uniquely
determine the number of edge modes at a given quasi-energy~\cite{Rudner13}. In particular, for FTIs it is possible to 
have anomalous edge states appearing at the boundaries
of the FBZ. FTIs can therefore realize topological systems where even though the Chern number
of the band is zero, yet equal number of chiral edge modes appear above and below it. 

When the laser frequency is larger than the band-width, 
conventional Chern insulators are realized for moderate laser amplitudes, where by conventional we mean
that there are edge states only at the center of FBZ, and the Chern number equals the number
of chiral edge modes. The anomalous edge states at the Floquet zone boundaries 
typically appear for resonant lasers where the resonance creates effective band-inversions~\cite{Lindner11}, 
with the anomalous edge states appearing at these band-inversion
points. Edge states at the FBZ boundaries oscillate in time at higher frequencies relative to those edge states at the center
of the FBZ~\cite{Erhai14,Erhai14b}. This leads to a situation where not all the edge states 
contribute equally to dc transport when the samples are connected to leads~\cite{Torres14}. 
  
With all these unusual properties, one has to have
a clear picture of how all the different edge modes, the ones appearing at the center of the FBZ, and
the anomalous ones appearing in the boundaries of the FBZ, affect measurable quantities. Thus
we need to explore how these edge modes are occupied, and the current densities carried by them.
Remarkably, despite the intense activity in the field, this study has not been done, and we plan to undertake it here
for a closed system in the absence of external dissipation.

We present results for a FTI realized by irradiating 
half-filled graphene in a cylindrical geometry with zig-zag edges, by
a circularly polarized laser. We assume the system is closed so that the occupation of all resulting Floquet quasi-energy
states is completely determined by the laser switch on protocol. 

We study the four different 
topological phases summarized in Fig.~\ref{fig1}
and labeled as $P_{1,2,3,4}$. Of these four phases,
one of them corresponds to an off-resonant high frequency laser
($P_1$), and the remaining ($P_{2,3,4}$) correspond to resonant low frequency lasers.
Moreover, of these four cases, two ($P_{1,2}$) 
are conventional Chern insulators in that edge states appear only at the center of the FBZ, while for the other two
phases ($P_{3,4}$), anomalous edge states appear at the boundaries of the FBZ. 

For the above phases we determine the occupation probability of 
the bulk and edge states following a laser quench. Moreover from the edge state population, we give simple Landauer based arguments to estimate the conductance
of the edge modes. In doing so we arrive at estimates that are consistent with a Kubo formalism computation of the dc Hall conductance of
a bulk system with no boundaries~\cite{Dehghani15a,Dehghani15b}. Thus even though the conductance is not $Ce^2/h$ for
resonant lasers due to nonequilibrium occupation of bands, we uncover a bulk-boundary correspondence that persists even in 
the nonequilibrium system, where the Hall response for a spatially extended system without edges is of the same magnitude as the transport
via edge states populated in a nonequilibrium way for precisely the same system but now with spatial boundaries.

In addition to edge state occupation, we also study the average current density, a quantity that can be locally measured using 
magnetometers such as SQUIDs.
We find that the nonequilibrium population following the laser quench 
breaks inversion symmetry and creates a net sheet of circulating current flowing on the cylinder. 
We show that since the individual eigenstates are chiral, such a current density profile
results in a charge imbalance between the top and bottom edges of the cylinder. 

In order to understand the symmetries
of the current density following the laser quench, we explore the symmetries of the current density carried by individual Floquet eigenstates, 
and in the process highlight
how even though the instantaneous Hamiltonian has no special symmetries other than particle-hole symmetry, the Floquet Hamiltonian, on 
averaging over one laser cycle, shows some additional emergent symmetries such as inversion symmetry. We discuss the role of these symmetries
on the current and charge densities generated by the laser quench. 

We now briefly discuss the relation between our work and existing literature. 
Our study is in a regime complementary to Ref.~\onlinecite{Torres14}
where a small sample in contact with leads was studied, and where the role of the anomalous edge modes is 
determined by how well they hybridize with lead states. In contrast our study is for larger systems and also closed 
systems such as those realized in ultra cold atomic gases~\cite{Esslinger14}. Our results are also relevant for
pump-probe spectroscopy of solid-state systems~\cite{Gedik13}, where time-resolved and angle resolved photoemission (ARPES) is a very effective way
of probing edge-state occupation probabilities.

Note that how edge states are occupied after a quench between two different static Hamiltonians 
with different topological invariants has been studied.~\cite{Bhaseen15,Galilo15}
In our work we study quench dynamics for a case where the final Hamiltonian after the quench is not static but is periodic in time. 
By virtue of this time-periodicity the 
edge state structure is far richer than in conventional TIs, leading to richer dynamics. 
Ref.~\onlinecite{Rigol14} studied dynamics in a similar system as
ours, however they focused only on the high frequency off-resonant case where the edge state structure is more
conventional.  Here in contrast we study both off-resonant and resonant laser frequencies, thus highlighting 
how the anomalous edge states are populated. In addition even for the off-resonant laser, our results are qualitatively
different from Ref.~\onlinecite{Rigol14} as our geometry, filling factor, and laser switch-on protocol results in a completely different
steady-state current density profile.

The paper is organized as follows. In Section~\ref{model}, we present the model and derive expressions for the
occupation probabilities and current densities. In Section~\ref{results} we present our results, while we
conclude in Section~\ref{conclu}, and give additional details in three appendices.

\section{Model} \label{model}

We consider graphene at half-filling in a cylindrical geometry with zig-zag edges that support edge states. 
The graphene sheet is irradiated by a circularly polarized and spatially uniform laser of
amplitude $A_0$ and frequency $\Omega$. 
Choosing $x$ to be the spatially uniform direction wrapping the cylinder, with $k_x$ being the momentum along
this direction, and labeling the sites along the
cylinder by $n_y=1 \ldots N_y$, where $N_y$ is even, the Hamiltonian of graphene without the laser is,
\begin{eqnarray}
&&H_G = -t_h\sum_{k_x,n_y=1\ldots N_y/2}\biggl[c^{\dagger}_{2n_y-1,k_x}c_{2n_y,k_x}\nonumber\\
&&\times \biggl(e^{-ik_x\delta_{1x}}+ e^{-ik_x\delta_{2x}}\biggr)
+h.c\biggr]\nonumber\\
&&+ \biggl[c^{\dagger}_{2n_y+1,k_x}c_{2n_y,k_x}+ h.c.\biggr]\biggl(1-\delta_{n_y=N_y/2}\biggr)\label{H1}.
\end{eqnarray}
Above odd and even sites are the $A$ and $B$ sub-lattices respectively, and the nearest-neighbor vectors measured from
the $B$ sub-lattice are, 
\begin{eqnarray}
\vec{\delta}_1=\frac{a}{2}\left(\sqrt{3},-1\right);\,\, \vec{\delta}_2=\frac{a}{2}\left(-\sqrt{3},-1\right);\,\, 
\vec{\delta}_3=a\left(0,1\right).
\end{eqnarray}
The laser enters through the replacement $c^{\dagger}_{\vec{r}'+\vec{r}}c_{\vec{r}'} \rightarrow c^{\dagger}_{\vec{r}'+\vec{r}}c_{\vec{r}'}
e^{-i\int_{\vec{r}'}^{\vec{r}'+\vec{r}}\vec{A}\cdot{d\vec{l}}}$. Thus
in the presence of a laser, the Hamiltonian gets modified to,
\begin{eqnarray}
&&H = -t_h\sum_{k_x,n_y=1\ldots N_y/2}
\biggl[c^{\dagger}_{2n_y-1,k_x}c_{2n_y,k_x}\nonumber\\
&&\times \biggl(e^{-ik_x\delta_{1x}-i \vec{A}\cdot\vec{\delta}_1}+ 
e^{-ik_x\delta_{2x}-i \vec{A}\cdot\vec{\delta}_{2}}\biggr)+h.c.\biggr]\nonumber\\
&&+ \biggl[c^{\dagger}_{2n_y+1,k_x}c_{2n_y,k_x}e^{-i\vec{A}\cdot\vec{\delta}_3}+ h.c.\biggr]\biggl(1-\delta_{n_y=N_y/2}\biggr),
\label{Hedge1}
\end{eqnarray}
where $\vec{A}= f(t)A_0\left[\cos(\Omega t),-\sin(\Omega t)\right]$ is the circularly polarized laser, and 
$f(t)$ is a function that determines how the
laser was switched on. In this paper we will study the effect of a sudden quench which
corresponds to $f(t) =\Theta(t)$, $\Theta(x)$ being the Heaviside
function. Physically this corresponds to time-evolving the ground state of graphene by the Hamiltonian $H(t>0^+)$.

Before the laser is switched on, the wavefunction corresponds to the
half-filled ground-state of graphene $|\Psi_{\rm in}\rangle$ which in Fock-space 
we write as,
\begin{eqnarray}
|\Psi_{\rm in}\rangle = \prod_{k_x,l= {\rm occ}}\epsilon^{\dagger}_{l,k_x}|0\rangle.
\end{eqnarray}
Above $l$ labels the exact eigenstates of graphene, there are $N_y$ of them for each $k_x$, and $l={\rm occ}$
implies the lowest $N_y/2$ occupied levels. These exact eigenstates can be expanded in the position
basis as,   
\begin{eqnarray}
\epsilon_{l,k_x}^{\dagger}=\sum_{n_y=1\ldots N_y}a_{k_x,l,n_y}c_{n_y,k_x}^{\dagger},
\end{eqnarray}
where $a_{k_x,l,n_y}$ are complex coefficients.

In the Heisenberg representation, the switching on of the laser implies the time-evolution,
\begin{eqnarray}
&&\frac{d}{dt}c_{n_y,k_x} = i\left[H(t), c_{n_y,k_x}(t)\right]\nonumber\\
&&=-i\sum_{n_y'}\biggl[h_{k_x}(t)\biggr]_{n_y,n_y'}c_{n_y',k_x}(t),
\end{eqnarray}
where we have denoted the full Hamiltonian as $H=\sum_{k_x,n_y,n_y'}c_{n_y,k_x}^{\dagger}\left[h_{k_x}\right]_{n_y,n_y'}c_{n_y',k_x}$.
The solution of the above equation is
\begin{eqnarray}
&&c_{n_y,k_x}(t) = \sum_{n_y'}\biggl[U_{k_x}(t,0)\biggr]_{n_y,n_y'}c_{n_y',k_x}(0)\label{ctimev},
\end{eqnarray}
where $U_{k_x}$ is an $N_y\times N_y$ unitary matrix representing the time-evolution operator,  
\begin{eqnarray}
i\frac{\partial}{\partial t}U_{k_x}(t,t')= H(t) U_{k_x}(t,t')\label{te1},
\end{eqnarray}
and obeys $U_{k_x}(t,t)=1$.

At times after the complete switch-on of the laser ($t,t'> 0^+$ for the quench) ,
\begin{eqnarray}
U_{k_x}(t,t') \!=\!\sum_{\alpha=1\ldots N_y}e^{-i\epsilon_{k_x\alpha}(t-t')}|\phi_{k_x,\alpha}(t)\rangle\langle \phi_{k_x,\alpha}(t')|,
\label{te3}
\end{eqnarray}
$\epsilon_{k_x\alpha}$ being the quasi-energies, while $|\phi_{k_x,\alpha}(t)\rangle$ are the time-periodic Floquet
quasi-modes~\cite{Sambe73}. In our representation these are $N_y$ component vectors whose components
we label as $\phi_{k_x,\alpha,n_y}$. Note the distinction between the time-periodic quasi-modes, and the
exact solution of the time-dependent Schr\"odinger equation $|\psi_{k_x,\alpha}(t)\rangle$,  where
the latter is obtained from the former by multiplication by a time-dependent phase,
\begin{eqnarray}
|\psi_{k_x,\alpha}(t)\rangle= e^{-i\epsilon_{k_x\alpha}t}|\phi_{k_x,\alpha}(t)\rangle.
\end{eqnarray}

We obtain the quasi-energies and quasi-modes using standard methods~\cite{Sambe73}. The time-periodicity of the
Floquet modes allows an expansion in Fourier components,
\begin{eqnarray}
|\phi_{k_x,\alpha}(t)\rangle = \sum_m e^{i m\Omega t}|\phi_{k_x,\alpha}^m\rangle\label{fexp}.
\end{eqnarray}
Eq.~\eqref{te1} implies that the Fourier components $\phi_{k_x,\alpha,n_y}^m$ obey,
\begin{eqnarray}
&&\sum_{m}\biggl[H^{n,m}+m\Omega \delta_{m,n}\biggr]|\phi_{k_x,\alpha}^m\rangle=\epsilon_{k_x\alpha}|\phi^n_{k_x,\alpha}\rangle,\\
&&H^{n,m}= \frac{\Omega}{2\pi}\int_0^{2\pi/\Omega}dte^{-i(n-m)\Omega t}H(t).
\end{eqnarray}
Thus the time-dependent problem of Eq.~\eqref{te1}, has been traded for a time-independent problem, albeit in an expanded Hilbert 
space due to the Fourier expansion.

In practice how many harmonics $|\phi^m\rangle$ need to be kept depends on the laser
parameters. Denoting the range of harmonics retained as $m=-M\ldots M$, we need to effectively solve
for the eigen-system of a $N_y(2M+1)\times N_y(2M+1)$ dimensional Hamiltonian.   
High frequency and low amplitudes usually require retaining fewer harmonics than low frequency and large amplitudes.
For the four cases studied by us, we find good numerical convergence for $M=6$ for phases $P_{1,2,3}$, and $M=12$ for the phase $P_4$.
Once the Fourier components $|\phi^m\rangle$ are known, the Floquet modes at any time can be obtained from 
Eq.~\eqref{fexp}, and the corresponding time-evolution operator can be determined from Eq.~\eqref{te3}.

A key physically relevant quantity entering in the expectation value of observables 
is the occupation probability $O_{\alpha}(k_x)$ of the Floquet eigenstates labeled by $k_x,\alpha$.
For the quench this is simply given by overlaps between the Floquet eigenstate at $t=0$, and the 
half-filled ground-state of graphene. To see this consider the simple case where initially only
a single mode of graphene labeled by $l$ is occupied. Thus the initial wave-function is
$|\psi_{k_x,\rm in}(0)\rangle = \epsilon_{l,k_x}^{\dagger}|0\rangle$. The quench implies, that from $t>0$, the state is 
\begin{eqnarray}
&&|\Psi_{k_x}(t)\rangle = U_{k_x}(t,0)|\psi_{k_x,\rm in}(0)\rangle\nonumber\\
&&=\sum_{\alpha}e^{-i\epsilon_{k_x\alpha}t}|\phi_{k_x,\alpha}(t)\rangle\langle \phi_{k_x,\alpha}(0)|\epsilon_{l,k_x}^{\dagger}|0\rangle.
\end{eqnarray} 
where in the last line we have used Eq.~\eqref{te3} for the time-evolution operator. What the above expression
implies is that the amplitude for being in the exact eigenstate of the
time-periodic Hamiltonian $|\psi_{k_x,\alpha}(t)\rangle=e^{-i\epsilon_{k_x\alpha}t}|\phi_{k_x,\alpha}(t)\rangle$ (which is the Floquet mode multiplied
by a phase), is a time-independent quantity and simply given by the overlap of the initial state and the exact eigenstate
at the time when the laser was switched on. We chose this time to be $t=0$. Thus the probability of being in the
exact eigenstate $k_x,\alpha$ is $|\langle \phi_{k_x,\alpha}(0)|\epsilon_{l,k_x}^{\dagger}|0\rangle|^2$

Accounting for the fact that initially not just one mode $l$, but many modes are occupied, the occupation
probability of the $\alpha$ quasi-energy level is simply obtained from summing over all the initially occupied states,
\begin{eqnarray}
&&O_{\alpha}(k_x) = \sum_{l={\rm occ}}|\langle \phi_{k_x,\alpha}(0)|\epsilon^{\dagger}_{l,k_x}|0\rangle|^2\nonumber\\
&&=\! \!\!\!\sum_{l={\rm occ}, n_y,n_y'}\!\!\!\biggl[\phi_{k_x,\alpha, n_y}^*(0)a_{k_x,l,n_y}\biggr]
\!\!\biggl[\phi_{k_x,\alpha, n_y'}(0)a^*_{k_x,l,n_y'}\biggr].
\label{Okxa}
\end{eqnarray}
One way to understand the meaning of these occupation probabilities is that since the final Hamiltonian is quadratic, it has 
many conserved quantities, which by definition do not evolve in time. $O_{\alpha}(k_x)$ should be viewed as these conserved
quantities. As shown further below, these are also the natural quantities entering in physical observables.
It is also useful to study the momentum averaged occupation of the Floquet levels,
\begin{eqnarray}
O_{\alpha}= \frac{1}{N_x}\sum_{k_x}O_{\alpha}(k_x)\label{Oa},
\end{eqnarray}
where $N_x$ is the number of points in the $\hat{x}$ direction.
 
We are interested in the current density operator as this directly measures the nature of the chiral eigenstates
of the periodically driven system. In order to define a current operator, we apply a weak vector potential $\vec{A}_{\rm pr}$,
and expand the Hamiltonian to leading order in it. Thus,
\begin{eqnarray}
&&H(t)\rightarrow H(t) -i\sum_{rr'ab}c_{r'+r,a}^{\dagger}h_{r'+r,r'}^{ab}(t)c_{r',b}\nonumber\\
&&\times \vec{r}\cdot \vec{A}_{\rm pr}(r'+\frac{r}{2}),
\end{eqnarray}
where $a,b$ is the graphene sublattice index.
For simplicity, let us say that the vector potential is spatially uniform and applied along the
$x$-direction, then $H(t) = H(A_{\rm pr}=0) + \hat{J}_x A_{\rm pr}$,
where $\hat{J}_x$ is the current operator in the $x$-direction,  
\begin{eqnarray}
\hat{J}_x= -i\sum_{r'rab}r_x c_{r'+r,a}^{\dagger}h_{r'+r,r'}^{ab}(t)c_{r',b}.
\end{eqnarray}
Because the system is spatially uniform along the $x$-direction, we perform a Fourier transform and write, 
\begin{eqnarray}
&&\hat{J}_x=- i\frac{1}{N_x}\sum_{k_{1x}k_{2x}ab,r,r'}c_{k_{1x},r_y'+r_y,a}^{\dagger}c_{k_{2x},r_y',b}r_x\nonumber\\
&&\times e^{i{k}_{2x}r'_x - ik_{1x}(r'_x+r_x)}h_{r'+r,r'}^{ab}(t).
\end{eqnarray}
Since $h^{ab}_{r'+r,r'}$ depends only on $r$, the sum on $r_x'$ gives $k_{1x}=k_{2x}$. Then, 
\begin{eqnarray}
&&\hat{J}_x
=\frac{1}{N_x}\sum_{k_{x},a,b,r_y',r_y}c_{k_{x},r_y'+r_y,a}^{\dagger}c_{k_{x},r_y',b}\nonumber\\
&&\times \partial_{k_x}\sum_{r_x}e^{- ik_{x}r_x}h_{r_x,r_y}^{ab}(t),\nonumber\\
&&=\frac{1}{2}\sum_{n_y=1\ldots N_y/2}\biggl(\hat{J}_{2n_y-1} + \hat{J}_{2n_y}\biggr),
\end{eqnarray}
where $\hat{J}_{n_y}$ is the current density at site $n_y$. Note that the current density in a unit-cell is the average of the
current density from the $A$ ($J_{2n_y-1}$) and $B$ sub-lattice ($J_{2n_y}$).
Using Eq.~\eqref{Hedge1}, current densities from the $A$ and $B$ sub-lattice are equal, and given by
\begin{eqnarray}
&&\hat{J}_{2n_y-1}
=\frac{t_h a}{N_x}\sqrt{3}\sum_{k_x}\biggl[c^{\dagger}_{2n_y-1,k_x}c_{2n_y,k_x}e^{-i\frac{A_0a}{2}\sin(\Omega t)}\nonumber\\
&&\times \sin\biggl(\frac{\sqrt{3}a}{2}\biggl\{k_x+A_0\cos(\Omega t)\biggr\}\biggr)+h.c.\biggr]\nonumber\\
&&= \hat{J}_{2 n_y}\label{curr1}.
\end{eqnarray}

It is convenient to expand the time-periodic matrix elements of the current operator in the Fourier basis,
\begin{eqnarray}
&&e^{i\frac{A_0a}{2}\sin(\Omega t)}\sin\biggl[\frac{\sqrt{3}a}{2}
\biggl\{k_x+A_0\cos(\Omega t)\biggr\}\biggr]\nonumber\\
&&= \sum_{m}e^{-im\Omega t}\tilde{J}_{-m}\left(A_0a\right)\sin\biggl(\frac{\sqrt{3}k_x a}{2}-\frac{m\pi}{3}\biggr),
\end{eqnarray}
where $\tilde{J}_m$ are the Bessel functions.
Thus the current density operator is
\begin{eqnarray}
&&\hat{J}_{2n_y-1}= \nonumber\\
&&\sqrt{3}\frac{t_h a}{N_x}\sum_{k_x,m}\biggl[\tilde{J}_{-m}\left(A_0a\right)\sin\biggl(\frac{\sqrt{3}k_x a}{2}-\frac{m\pi}{3}\biggr)\biggr]
\nonumber\\
&&\times 
\biggl[\biggl(e^{i m\Omega t}c^{\dagger}_{2n_y-1,k_x}c_{2n_y,k_x}+h.c\biggr)\biggr]\label{jac}.
\end{eqnarray}
It is interesting to note that if one retains only the $m=0$ harmonic, the current density operator is the same as that for
the undriven case, but with the effective hopping amplitude 
$t_h$ renormalized to $t_h \tilde{J}_0(A_0a)$ by the laser. For non-zero $m$, the above expression for the current operator highlights that the
electron tunneling between neighboring sites can be accompanied by $m$-photon absorption or emission processes, with 
$t_h \tilde{J}_m(A_0a)$ controlling the amplitude of such processes. 

\begin{figure}
\includegraphics[height=9cm,width=9cm,keepaspectratio]{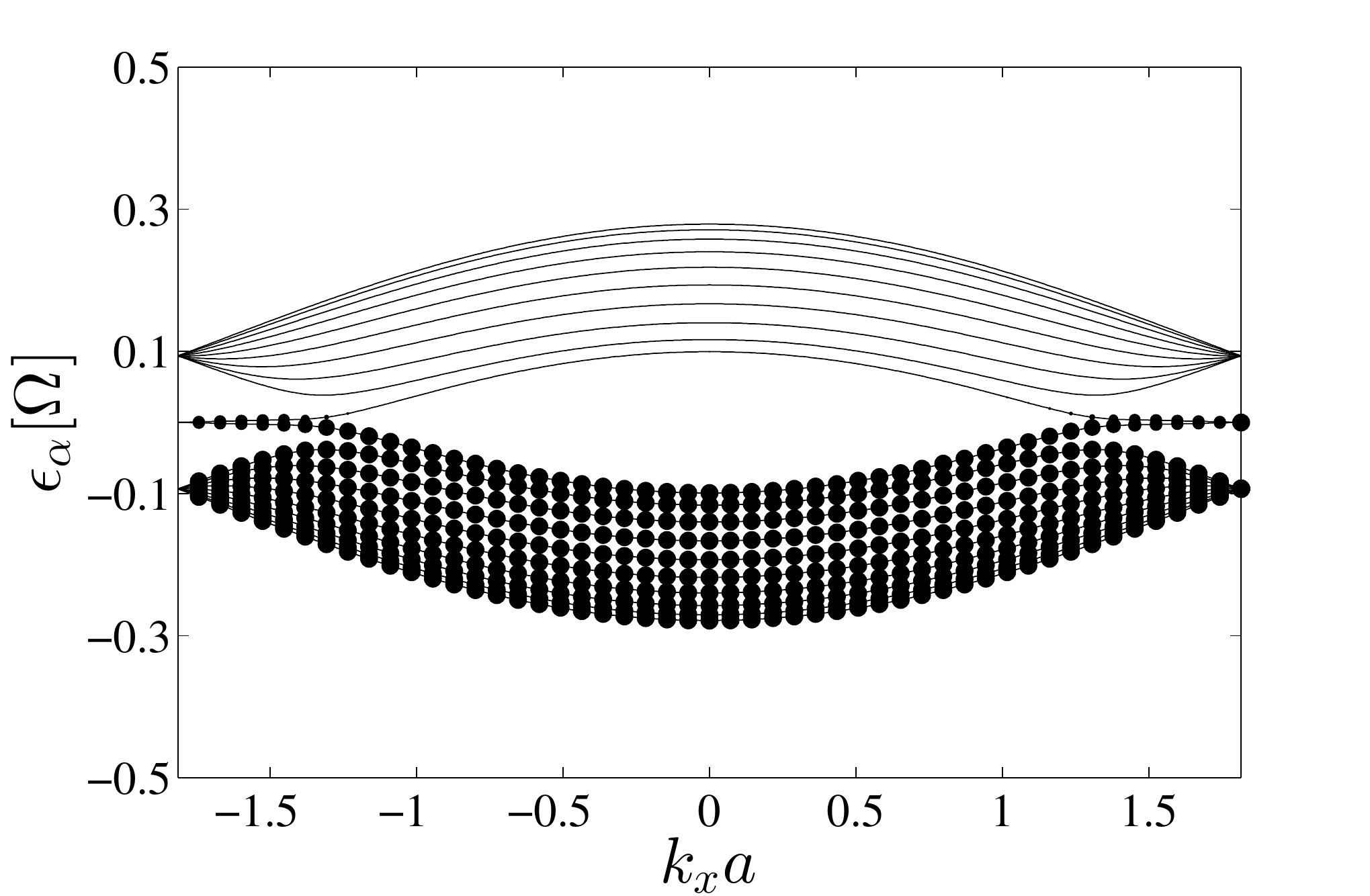}\\
\caption{
Spectrum and occupation probabilities due to a quench for the case $P_1$ where
$A_0a=0.5,\Omega=10t_h$, and the Chern number is $C=1$. The system
supports a pair of chiral edge modes at the center of the FBZ. The area of the circles are proportional to the occupation probability
$O_{\alpha}(k_x)$.
}
\label{fig2}
\end{figure}

\begin{figure}
\includegraphics[height=9cm,width=9cm,keepaspectratio]{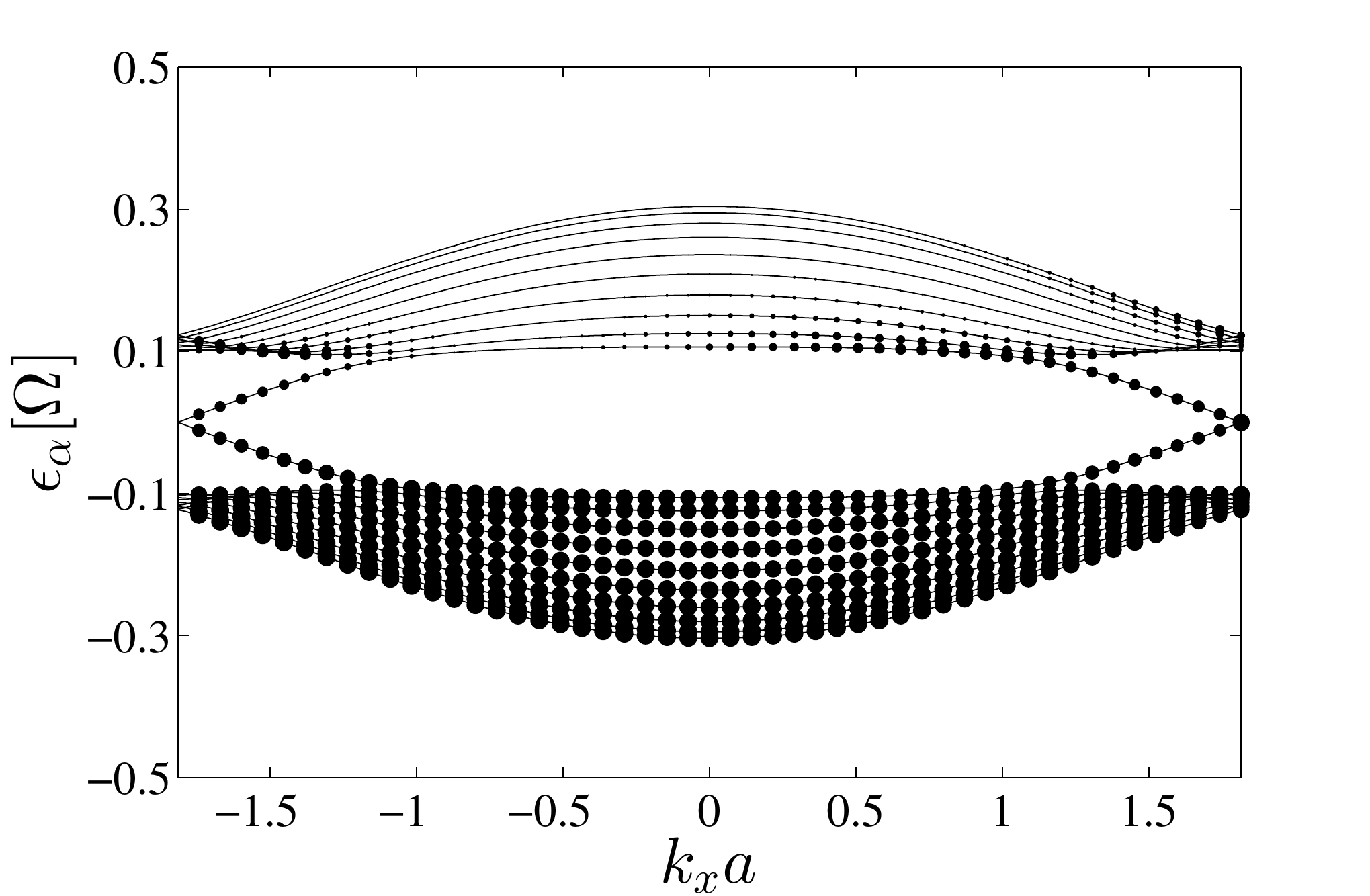}\\
\caption{
Spectrum and occupation probabilities due to a quench for the case $P_2$ where $A_0a=1.5,\Omega=5t_h$, 
and the Chern number is $C=1$. The system
supports a pair of chiral edge modes at the center of the FBZ. The area of the circles are proportional to the occupation probability
$O_{\alpha}(k_x)$. 
}
\label{fig3}
\end{figure}

The expectation value of the current density operator at a time $t$ after the quench is
\begin{eqnarray}
&&J_{2n_y-1}(t)= \langle \Psi_{\rm in}|{\cal \tilde{T}}e^{i\int_0^t dt' H(t')}\hat{J}_{2n_y-1}{\cal T}e^{-i\int_0^t dt' H(t')}|\Psi_{\rm in}\rangle\nonumber\\
&&= \sqrt{3}\frac{t_h a}{N_x}\sum_{k_x,m}\biggl[\tilde{J}_{-m}\left(A_0a\right)\sin\biggl(\frac{\sqrt{3}k_x a}{2}-\frac{m\pi}{3}\biggr)\biggr]\nonumber\\
&&\times 
\biggl[\langle \Psi_{\rm in}|\biggl(e^{i m\Omega t}c^{\dagger}_{2n_y-1,k_x}(t)c_{2n_y,k_x}(t)+h.c\biggr)|\Psi_{\rm in}\rangle\biggr].
\label{jac1}
\end{eqnarray}
Above ${\cal T}, {\cal \tilde{T}}$ are the time and anti-time ordering operators respectively, and the time-dependent behavior of $c_{2n_y,k_x}(t)$
are obtained from Eq.~\eqref{ctimev}.

In Eq.~\eqref{jac1} we need to evaluate expectation values of the 
kind $n_{j_1,j_2,k_x}(t) = \langle \Psi_{\rm in}|c_{j_1,k_x}^{\dagger}(t)c_{j_2,k_x}(t)|\Psi_{\rm in}\rangle$,
which using the time-evolution operator may be written as,
\begin{eqnarray}
&&n_{j_1,j_2,k_x}(t) = \langle \Psi_{\rm in}|c_{j_1,k_x}^{\dagger}(t)c_{j_2,k_x}(t)|\Psi_{\rm in}\rangle\nonumber\\
&&=\sum_{j'j''}\biggl[U_{k_x}(t,0)\biggr]_{j_2j'}\biggl[U_{k_x}(0,t)\biggr]_{j''j_1} \nonumber\\
&&\times \langle \Psi_{\rm in}| c_{j'',k_x}^{\dagger}(0) 
c_{j',k_x}(0)|\Psi_{\rm in}\rangle\nonumber\\
&&=\sum_{j'j'', l={\rm occ}}
\biggl[U_{k_x}(t,0)\biggr]_{j_2j'}\biggl[U_{k_x}(0,t)\biggr]_{j''j_1} a_{k_x,l,j'}a^*_{k_x,l,j''}\nonumber\\
&&= \sum_{j',j'',\alpha\beta,l={\rm occ}} e^{-i\epsilon_{k_x\alpha} t + i\epsilon_{k_x\beta} t}\phi_{k_x,\alpha,j_2}(t) \phi_{k_x,\alpha,j'}^*(0)\nonumber\\
&&\times \phi_{k_x ,\beta,j''}(0)\phi^*_{k_x,\beta,j_1}(t)a_{k_x,l,j'}a^*_{k_x,l,j''}\label{nij}.
\end{eqnarray}
At long times, we need only keep $\alpha=\beta$ terms, as the $\alpha \neq \beta $ terms oscillate in time with different frequencies
for the different momenta $k_x$. Thus on summing over 
$k_x$ the $\alpha\neq \beta $ terms vanish as a power-law due to dephasing.
Thus at long times, after the dephasing has set in, the current density is given by the ``diagonal ensemble'' corresponding to keeping only $\alpha=\beta$, 
\begin{eqnarray}
&&J_{2n_y-1}(t\rightarrow \infty)= \nonumber\\
&&\sqrt{3}\frac{t_h a}{N_x}\sum_{k_x,m,j',j'',\alpha,l={\rm occ}}
\biggl[e^{i m\Omega t}\phi_{k_x,\alpha,2n_y}(t)\phi^*_{k_x,\alpha,2n_y-1}(t)\nonumber\\
&&\times  \phi_{k_x,\alpha, j'}^*(0)\phi_{k_x,\alpha, j''}(0)a_{k_x,l,j'}a^*_{k_x,l,j''}+h.c.\biggr]\nonumber\\
&&\times \biggl[\tilde{J}_{-m}\left(A_0a\right)\sin\biggl(\frac{\sqrt{3}k_x a}{2}-\frac{m\pi}{3}\biggr)\biggr].
\end{eqnarray}
This result is still oscillatory over the period of the laser on account of the time periodicity of the Floquet modes. 
Expanding the Floquet modes in their Fourier basis $\phi(t)=\sum_me^{i m \Omega t}\phi^m$, and time averaging over one cycle of the
laser, we find, the quench current density to be,
\begin{eqnarray}
J_{n_y}(t\rightarrow \infty)= \frac{1}{N_x}\sum_{k_x,\alpha={1\ldots N_y} }O_{\alpha}(k_x)j_{\alpha,n_y}(k_x)\label{currq},
\end{eqnarray}
where $j_{\alpha,n_y}(k_x)$ is the current density carried by an individual Floquet eigenstate labeled by $\alpha, k_x$ and
time averaged over a laser cycle,
\begin{eqnarray}
&&j_{\alpha,2n_y-1}(k_x)=\nonumber\\
&&\sqrt{3}t_h a\sum_{m,n}\biggl[\tilde{J}_{-m}\left(A_0a\right)\sin\biggl(\frac{\sqrt{3}k_x a}{2}-\frac{m\pi}{3}\biggr)\biggr]
\nonumber\\
&&\times 2{\rm Re}\biggl[\phi_{k_x,\alpha,2n_y}^n\biggl(\phi_{k_x,\alpha,2n_y-1}^{n+m}\biggr)^* \biggr],\label{curres}\\
&& j_{\alpha,2n_y}(k_x)= j_{\alpha,2n_y-1}(k_x).\label{curres2a}
\end{eqnarray}

Above we see the key role played by the occupation $O_{\alpha}(k_x)$ as the 
current density at site $n_y$ is the current density $j_{\alpha,n_y}(k_x)$  
from all Floquet eigenstates ($\alpha,k_x$) weighted by the occupation of the states. 
In what follows we will not only discuss the quench current density defined in Eq.~\eqref{currq},  but also the current density of a single
Floquet eigenstate $\alpha$ when the occupation is the same for all $k_x$. This is defined as,
\begin{eqnarray}
J_{\alpha,n_y}= \frac{1}{N_x}\sum_{k_x}j_{\alpha,n_y}(k_x), \label{curres2}
\end{eqnarray}
and is simply the momentum average of the current density operator of a Floquet eigenstate.

Note that the current density is not what is directly measured in transport such as Hall response. For the latter, 
a proper Kubo formula or Landauer formalism approach needs to be employed.
Employing Kubo formalism, one finds that the Hall current is determined by topological properties such as the time-averaged Berry
curvature, but now weighted by the occupation probabilities of the quasi-energy bands~\cite{Dehghani15a}. 
The average current density on the other hand is far more sensitive to microscopic details, and can be probed using other methods
such as sensitive magnetometers like SQUIDs that respond to the local magnetization generated by 
local currents~\cite{Bluhm09,Shibata15}.

\section{Results}\label{results}
The laser frequency  and amplitude can be used to drive a series of topological phase transitions, and in this paper we focus 
on the four topological phases summarized in Fig.~\ref{fig1}. Of these four phases, $P_1$ corresponds to a high-frequency
off-resonant laser where the laser frequency is larger than the bandwidth of graphene ($ \Omega > 6 t_h$). The 
other three phases correspond to low-frequency resonant lasers ($ \Omega < 6 t_h$).
We first discuss the occupation probabilities for these four cases separately below, followed by a discussion of the
current densities.

\subsection{Occupation probability and bulk-boundary correspondence
in transport}
The phase $P_1$ corresponds to an off-resonant laser with parameters
$A_0a=0.5,\Omega = 10 t_h$ and Chern number $C=1$.  The quasi-energies for this case are shown in Fig.~\ref{fig2}, and
include a pair of
chiral edge modes at the center of the FBZ. Thus for this case, the Chern number equals the number of chiral edge modes.
A key quantity is the occupation probabilities of the edge and bulk modes. For a quench switch-on protocol these
are given by Eq.~\eqref{Okxa}, and are quite simply determined by the 
overlap of the Floquet modes at $t=0$ with the occupied states of
graphene. 

The occupation probabilities of the Floquet levels for $P_1$ are indicated by circles in Fig.~\ref{fig2},
with the area of the circles proportional to the occupation $O_{\alpha}(k_x)$. The effect of the quench on the 
occupation can also be summarized by simply taking the momentum average as defined in Eq.~\eqref{Oa},
and as plotted in the top left panel of Fig.~\ref{fig6}. 
These figures 
show that for the off-resonant laser, to a good degree, only the lower Floquet band is occupied even though the 
laser was switched on as a quench. Fig.~\ref{fig6} in fact shows that the distribution function for $P_1$ looks like a 
zero temperature Fermi-function of a half-filled state. 

For all the phases studied here, one finds the following symmetry between the occupation probabilities for quasi-levels
with quasi-energies of the opposite sign,
\begin{eqnarray}
O_{\alpha}(k_x) + O_{N_{y} - \alpha + 1}(k_x)= 1.\label{DistFuncCompl}
\end{eqnarray}
We have proven the above relation in Appendix~\ref{app1}, where we show that it arises as a 
consequence of the particle-hole symmetry of the Hamiltonian before the quench, and the Floquet Hamiltonian.
The momentum average of the above expression is $O_{\alpha} + O_{N_{y} - \alpha + 1}= 1$, a behavior which is clearly reflected 
in all the panels of Fig.~\ref{fig6}.

In Ref.~\onlinecite{Dehghani15a} the effect of a quench switch-on of the laser was studied on bulk graphene, where there are no edge modes.
Several topological phases were studied, 
and the Hall conductance from a Kubo formula approach was computed. 
For the $P_1$ phase, the Hall conductance was found to
show a value close to the maximum value of $e^2/h$. This result is consistent
with our observation here that for the off-resonant laser, the quench occupation is very close
to that of an ideal half-filled Floquet band. 

The results of Ref.~\onlinecite{Dehghani15a} for an extended system with no boundaries,
and our results here for a finite system which hosts edge states show 
signatures of the bulk-boundary correspondence of TIs, where Hall response
in a bulk system can alternately be described in terms of transport by chiral edge states when a
infinitesimal chemical potential difference is applied to them.
This is because we find here that for $P_1$, the edge states in
the center of the FBZ survive the quench, and are occupied with a very low effective temperature. 
Thus this pair will contribute to a conductance of ${\cal O}(e^2/h)$ within a Landauer formalism
that assumes there is no inelastic scattering. Any deviations from this value
is due to the small, albeit non-zero excitations of the bulk 
states which have the opposite chirality to the edge states (see further discussion below).
The Hall response in closed systems without leads
can be measured experimentally in cold-atomic gases along the lines of  Ref.~\onlinecite{Esslinger14}
where an application of an external potential gradient leads to a transverse drift of atoms due to
a non-zero Chern number of the atomic bands. 

We make similar observations for the phase $P_2$
which now corresponds to a resonant laser where
$A_0a=1.5, \Omega = 5 t_h, C=1$. 
The spectrum is shown in Fig.~\ref{fig3}. Thus this phase is similar to phase $P_1$
in being like a conventional 
Chern insulator where the Chern number equals the number of chiral edge modes. 
An interesting observation is the asymmetry in $k_x$, i.e., $O_{\alpha}(k_x)\neq O_{\alpha}(-k_x)$. This exists even for $P_1$,
but is less visible there. 
The asymmetry in the occupation was also noticed in Ref.~\onlinecite{Dehghani14} where the effect of a 
quench switch-on of the laser was studied on a bulk system (with no edge modes). 
The asymmetry arises because the laser breaks 
inversion symmetry. For our case in particular, at the switch-on time $t=0$, the laser is pointed entirely along the $x$-direction,
thus breaking the inversion symmetry in $\hat{x}$. 

Both  the $O_{\alpha}(k_x)$ in Fig.~\ref{fig3}, as well as its momentum average 
in the top right panel of Fig.~\ref{fig6}  show that the $P_2$
case corresponds to a slightly higher effective temperature in comparison to
phase $P_1$,  with both lower ($\epsilon<0$) and upper ($\epsilon>0$) 
edge modes getting occupied, and a larger fraction of
the bulk states being occupied. Yet the bulk excitation density is still quite low like $P_1$.
A bulk Kubo formula computation for the dc Hall conductance in a spatially extended system for this case revealed~\cite{Dehghani15a} a 
result of ${\cal O}(e^2/h)$, consistent with the 
low excitation density of bulk states generated by the quench even for this phase. We would arrive at the same
conclusion if we were to alternately attribute the entire Hall response as due to the chiral edge-state studied here,
where the bulk excitations degrade the maximum possible value by a small amount.
 
While it is simple to understand why a pair of counter-propagating edge states at zero effective temperature (such
as those encountered so far) will give a linear response conductance of ${\cal O}(e^2/h)$,
we now briefly explain why a pair of counter-propagating edge states at infinite effective temperature will give
zero contribution to dc transport. If we think of the dc transport as a linear response to a small chemical potential
difference, a net current flows because the population of say the left mover is increased slightly over the right mover.
If now the pair were such that all states were uniformly occupied, a small voltage bias will not change the
net occupation between left and right movers, leading to zero conductance. 
This simple picture will come in handy when understanding the bulk-boundary correspondence in the phases $P_{3,4}$ below.

\begin{figure}
\includegraphics[height=9cm,width=9cm,keepaspectratio]{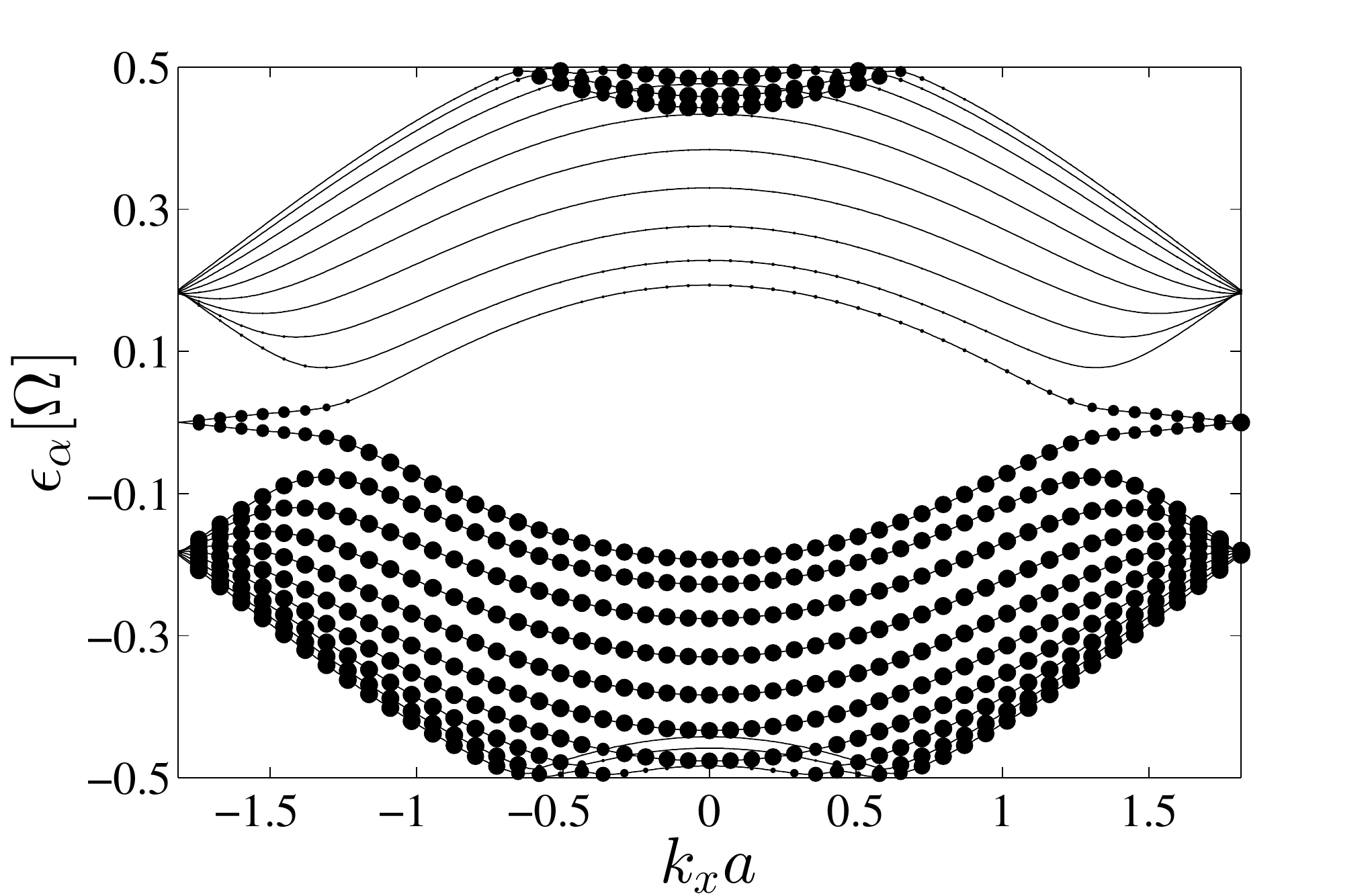}\\
\caption{
Spectrum and occupation probabilities due to a quench for the case $P_3$ where 
$A_0a=0.5,\Omega=5t_h$, and the Chern number is $C=3$. The system
supports a pair of chiral edge modes at the center of the FBZ, and two pairs of chiral edge modes on the
Floquet zone boundaries (see Fig.~\ref{fig1} and Fig.~\ref{fig8}). 
The area of the circles are proportional to the occupation probability
$O_{\alpha}(k_x)$.
}
\label{fig4}
\end{figure}

The phase $P_3$ is also a resonant laser corresponding to
$A_0a=0.5, \Omega = 5 t_h, C=3$, but it is very different from the resonant case $P_2$ discussed above. 
The spectrum for $P_3$ is shown in Fig.~\ref{fig4}, and reveals the unusual properties of the 
Floquet Chern insulator, where anomalous edge states
appear at the Floquet zone boundaries. The chiralities of these edge modes are shown schematically in Fig.~\ref{fig1},
and explicitly via the current densities in Fig.~\ref{fig8}.
Thus for this case the Chern number now equals the difference between the number of chiral
edge modes above and below the band, where there are two right movers above and one left mover below the 
band on one of the two spatial boundaries.  A clear signature of the laser 
resonance is seen in both Fig.~\ref{fig4} and the lower left panel of Fig.~\ref{fig6}. 
The resonance shows up as a selective depletion of the lower Floquet band, and the corresponding 
selective occupation of the upper band.
Note that the points in $k_x$ where the occupation changes suddenly due to the resonance condition, are 
also the points in $k_x$ at which anomalous edge states appear. This is because the laser resonance effectively produces a band
crossing at $|k_x|a\sim 0.5$ in Fig.~\ref{fig4}. This band crossing is accompanied by a change in the Chern number and a corresponding
change in the number of edge modes. 

Thus phase $P_3$ is special in that in the same phase one has edge modes that arise due to
off-resonant and resonant processes. 
The pair of edge modes located at
$\epsilon=0$ have the same origin as in the high frequency laser (phase $P_1$) as they arise due to off-resonant virtual processes,
while the two pairs of anomalous edge modes arise due to resonant processes. This is also reflected in the fact that the
edge modes at $\epsilon=0$ are occupied at a very low effective temperature (see the sharp step at $\alpha=N_y/2$ in
lower left panel of Fig.~\ref{fig6}), while the anomalous edge modes are at a much higher effective temperature. 
Note that anomalous edge modes are not so clearly visible in Fig.~\ref{fig4} as the quasi-energy gap in which they
live are rather small. However the corresponding current densities carried by them is shown in Fig.~\ref{fig8} and indeed
show the current density to be localized at the boundary. 

Interestingly for a quench in a bulk system, the case $P_3$ showed a Hall conductance of 
approximately $e^2/h$ in Ref.~\onlinecite{Dehghani15a}. This is  
a far deviation of $3 e^2/h$ for the Hall conductance 
if only the lower Floquet band was fully occupied. 
This came about because in the bulk computation of Ref.~\onlinecite{Dehghani15a}, and as can also be seen here in Fig.~\ref{fig4},
the resonance significantly populates portions of the upper Floquet band. Since the upper Floquet band
has the opposite Berry curvature to the lower one, it reduced the Hall conductance to almost
$1/3$ of its maximum value of $3 e^2/h$ in Ref.~\onlinecite{Dehghani15a}.

As a signature of the bulk-boundary correspondence
in topological systems, a dc Hall conductance of 
$\sim 1 e^2/h$ is consistent with our observation here that for the phase $P_3$, of the 3 pairs of edge states likely
to participate in transport, 2 of them, in particular the ones that reside at the boundaries of the FBZ are at a much higher effective temperature
as they arise due to resonant processes.
Thus these two pairs contribute relatively little to dc transport. Most of the dc transport in an edge state picture
comes from the off-resonant pair of edge states located at the center of the FBZ. Any further deviations from $1 e^2/h$
is due to residual bulk excitations that have the opposite chirality to the edge-state.

\begin{figure}
\includegraphics[height=9cm,width=9cm,keepaspectratio]{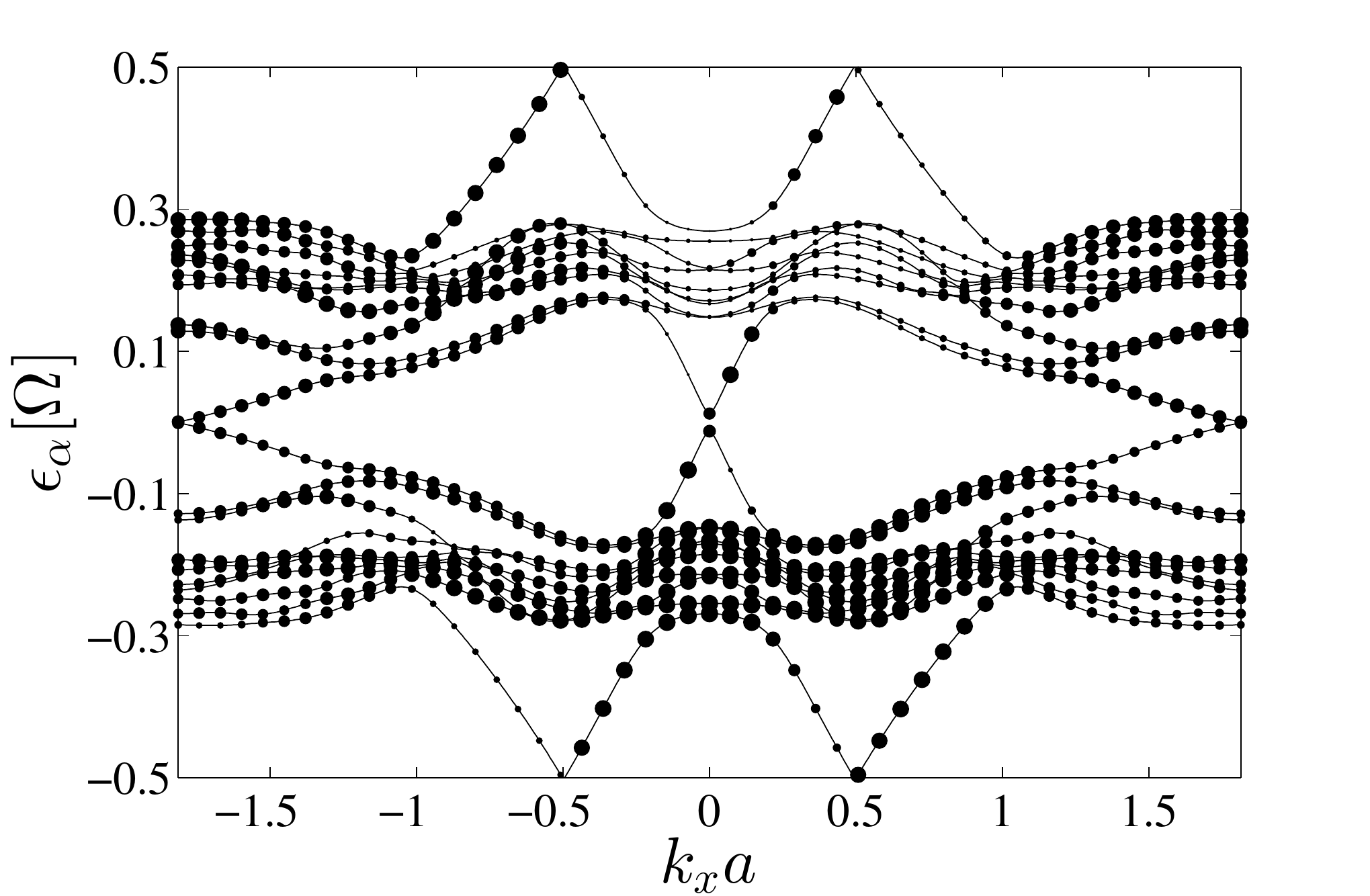}\\
\caption{
Spectrum and occupation probabilities due to a quench for the case $P_4$ where
$A_0a=10,\Omega=0.5t_h$, and the Chern number is $C=0$. The system
supports two pairs of chiral edge modes at the center of the FBZ, and two pairs of chiral edge modes on the
Floquet zone boundaries (see Fig.~\ref{fig1} and Fig.~\ref{fig8}). 
The area of the circles are proportional to the occupation probability $O_{\alpha}(k_x)$.
}
\label{fig5}
\end{figure}

The third resonant case corresponds to phase $P_4$ with laser
parameters $A_0a =10, \Omega=0.5 t_h,C=0 $. As the spectrum in Fig.~\ref{fig5} shows, 
this case also highlights a peculiarity of Floquet Chern insulators
in that it is possible to have bands with zero Chern number, and yet 
topological edge states appear above and below the quasi-band. For this case there are 4 pairs of edge states with the same chirality, 
with two of these residing
above the Floquet band, and two residing below the Floquet band. 
Lower right panel of Fig.~\ref{fig6} shows that $P_4$ is like an infinite temperature state as 
the occupation probabilities
of all the levels are almost the same. This is not surprising given that the laser frequency is much smaller 
than the band-width, leading to
many pockets of resonances. These pockets are not sharp like in phase $P_3$, but get smoothened out due to the large laser 
amplitude that increases the matrix elements for multi-photon processes.

The dc and optical Hall conductance for a bulk system for the same laser parameters as $P_4$ 
was studied in Ref.~\onlinecite{Dehghani15b}. It was found that a small albeit non-zero Hall conductance is possible for a 
laser quench, even though
an ideal occupation of the Floquet bands would lead to a zero Hall conductance. This non-zero conductance comes about because of the 
nonequilibrium occupation of the Floquet bands, each of which have a net chirality (see further discussion below) 
leading to a non-zero Hall
response. The magnitude of the Hall response is much smaller than $e^2/h$, and consistent with all 4 pairs of edge modes being 
occupied at an infinite effective temperature.
This qualitative similarity between the results in this paper
and the bulk computation of Refs.~\onlinecite{Dehghani15a,Dehghani15b} based on
the Kubo formalism, is a signature of the bulk-boundary correspondence that 
exists in topological systems, and appears to persist even out of equilibrium. 

\begin{figure}
\includegraphics[height=9cm,width=9cm,keepaspectratio]{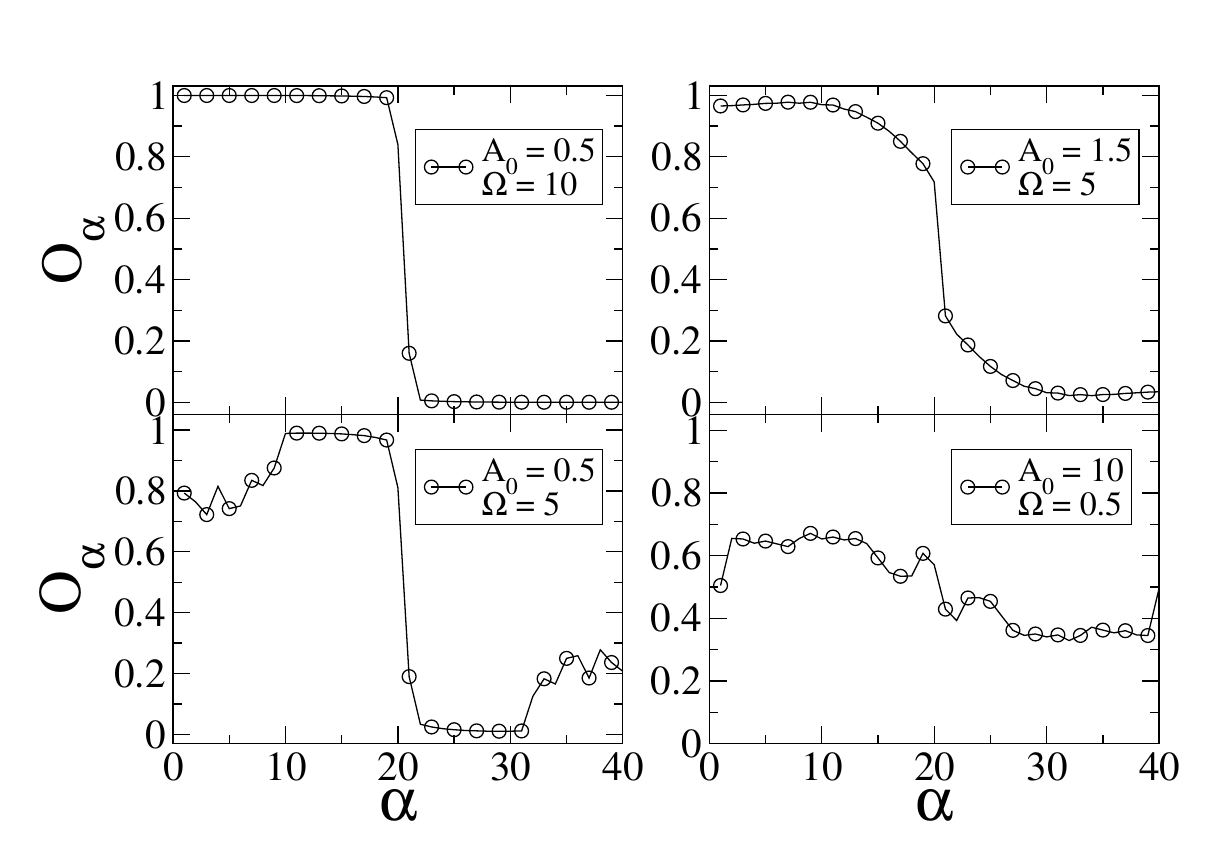}\\
\caption{
Quench occupation probabilities of the Floquet quasi-energy levels averaged over the momenta $k_x$. 
Clockwise from top left, phases $P_1$, $P_2$, $P_4$ and $P_3$ (see Fig.~\ref{fig1}). 
$P_4$ has an almost infinite effective temperature, while $P_1$ has an almost zero effective temperature. 
}
\label{fig6}
\end{figure}

\begin{figure}
\includegraphics[height=9cm,width=8.8cm,keepaspectratio]{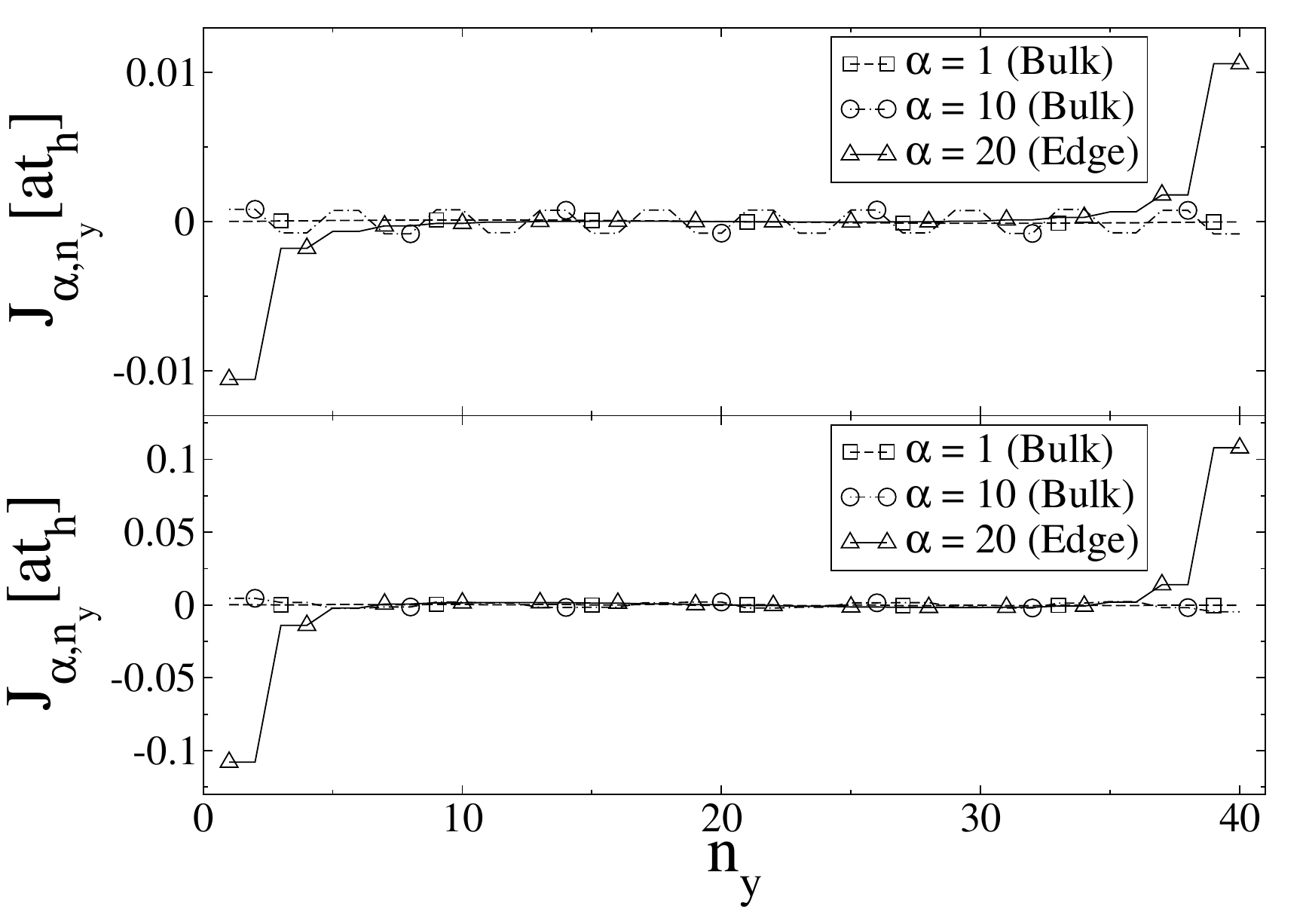}\\
\caption{Upper panel: Case $P_1$ where $A_0a=0.5,\Omega=10t_h,C=1$. Lower-panel: Case $P_2$ where $A_0a=1.5,\Omega=5t_h,C=1$.
Both correspond to strip width $N_y=40$.
Current densities of three exact Floquet eigenstates: one from the Floquet band edge ($\alpha=1$), another from
the center of the Floquet band ($\alpha=N_y/4=10$) and a third from the edge state located at the center of the 
FBZ ($\alpha=N_y/2=20$). The current
densities from the two bulk states $\alpha=1,10$ are much smaller than that from the edge states, and are also of the
opposite chirality from that of the edge state. 
}
\label{fig7}
\end{figure}

\subsection{Current density}

Finally we turn to the question of the current densities, a key quantity that directly probes the chiral nature of the system. 
$J_{\alpha,n_y}$,
defined in Eq.~\eqref{curres2}, is the current density of the $\alpha$ Floquet level given that all $k_x$ states are
equally occupied. In contrast, the quench current density given 
in Eq.~\eqref{currq} is the current density of the Floquet eigenstates, but weighted by
the occupation probabilities $O_{\alpha}(k_x)$ of these states. 
We first discuss the current densities of the Floquet eigenstates before we turn to the quench current density
which has contributions from all Floquet eigenstates. 

It is first useful to make some general observations. For a fully occupied band,
the current density is zero. This manifests in many ways, for example the two bulk-bands have opposite 
Chern number~\cite{Dehghani15a} so 
that when both bands are fully occupied, there is no Hall response. For our system with edges, this implies
\begin{eqnarray}
\sum_{\alpha=1\ldots N_y} J_{\alpha,n_y}=0.
\end{eqnarray} 

An important point to note is that even though the instantaneous $H(t)$
has no particular symmetry other than particle-hole symmetry, the Floquet Hamiltonian shows additional symmetries such as inversion symmetry 
when the Floquet modes are averaged over one cycle of the laser (see Appendices~\ref{app1},~\ref{app2}). 
This is also seen by noting that in the high-frequency limit, 
a Magnus expansion of the Floquet Hamiltonian yields the Haldane model with
particle-hole symmetry, inversion symmetry, but broken time-reversal symmetry.~\cite{Oka09,Kitagawa10}
As shown in Appendix~\ref{app3}, a consequence of these symmetries is that the current density carried
by a Floquet eigenstate time-averaged over a laser cycle is 
exactly anti-symmetric in position,
\begin{eqnarray}
J_{\alpha,n_y}=-J_{\alpha, N_y-n_y+1}.\label{CurrentInv}
\end{eqnarray}
Furthermore, there exists an exact symmetry between current densities from lower 
($1\leq \alpha \leq N_y/2$) and  upper ($ 1+N_y/2\leq  \alpha \leq N_y $) Floquet bands,
\begin{eqnarray}
J_{\alpha,n_y} = J_{N_y-\alpha+1,n_y}. \label{CurrentChiral} 
\end{eqnarray}

\begin{figure}
\includegraphics[height=9cm,width=8.7cm,keepaspectratio]{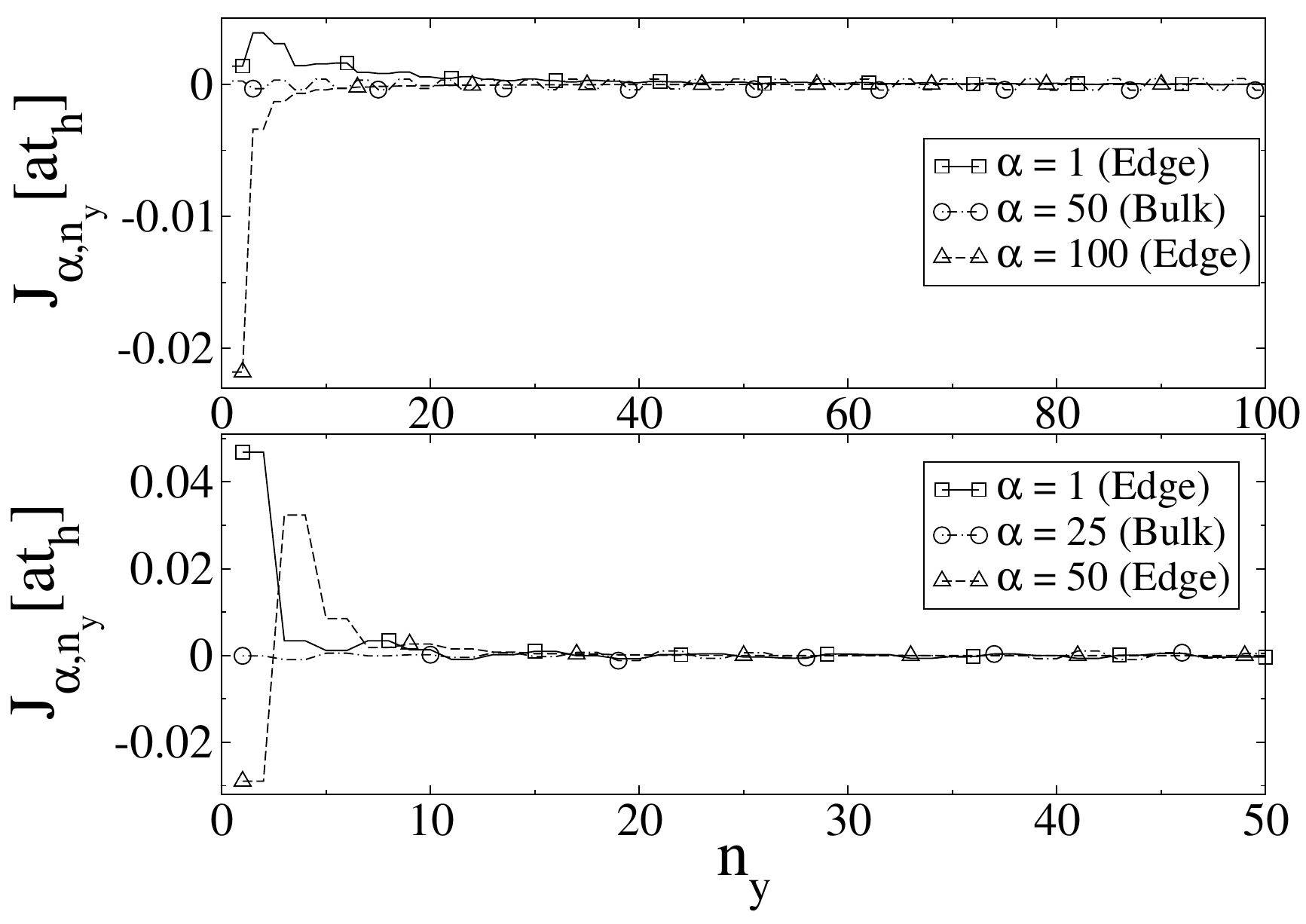}\\
\caption{Upper panel: Case $P_3$ where $A_0a=0.5,\Omega=5t_h,C=3$ with strip width $N_y=200$. 
Lower-panel: Case $P_4$ where $A_0a=10,\Omega=0.5t_h,C=0$
with strip width $N_y=100$. For each panel, 
current densities of three exact Floquet eigenstates, one state from the Floquet band edge ($\alpha=1$), another from
the Floquet band center ($\alpha=N_y/4=50 (P_3), 25 (P_4)$) and a third from the edge state(s) located at the center of the 
FBZ ($\alpha=N_y/2=100 (P_3), 50 (P_4)$) are shown. 
The current density from the bulk state $\alpha=N_y/4$ is much smaller than that from the edge states 
($\alpha=1,N_y/2$). The current density 
over only half the strip has been plotted as for the other half the current is anti-symmetric to the first half.
}
\label{fig8}
\end{figure}

The implications of this for the Floquet states is that for an exactly half-filled Floquet band, the current density
vanishes,
\begin{eqnarray}
\sum_{\alpha=1\ldots N_y/2} J_{\alpha,n_y}=0.
\end{eqnarray} 

\begin{figure}
\includegraphics[height=9cm,width=8.8cm,keepaspectratio]{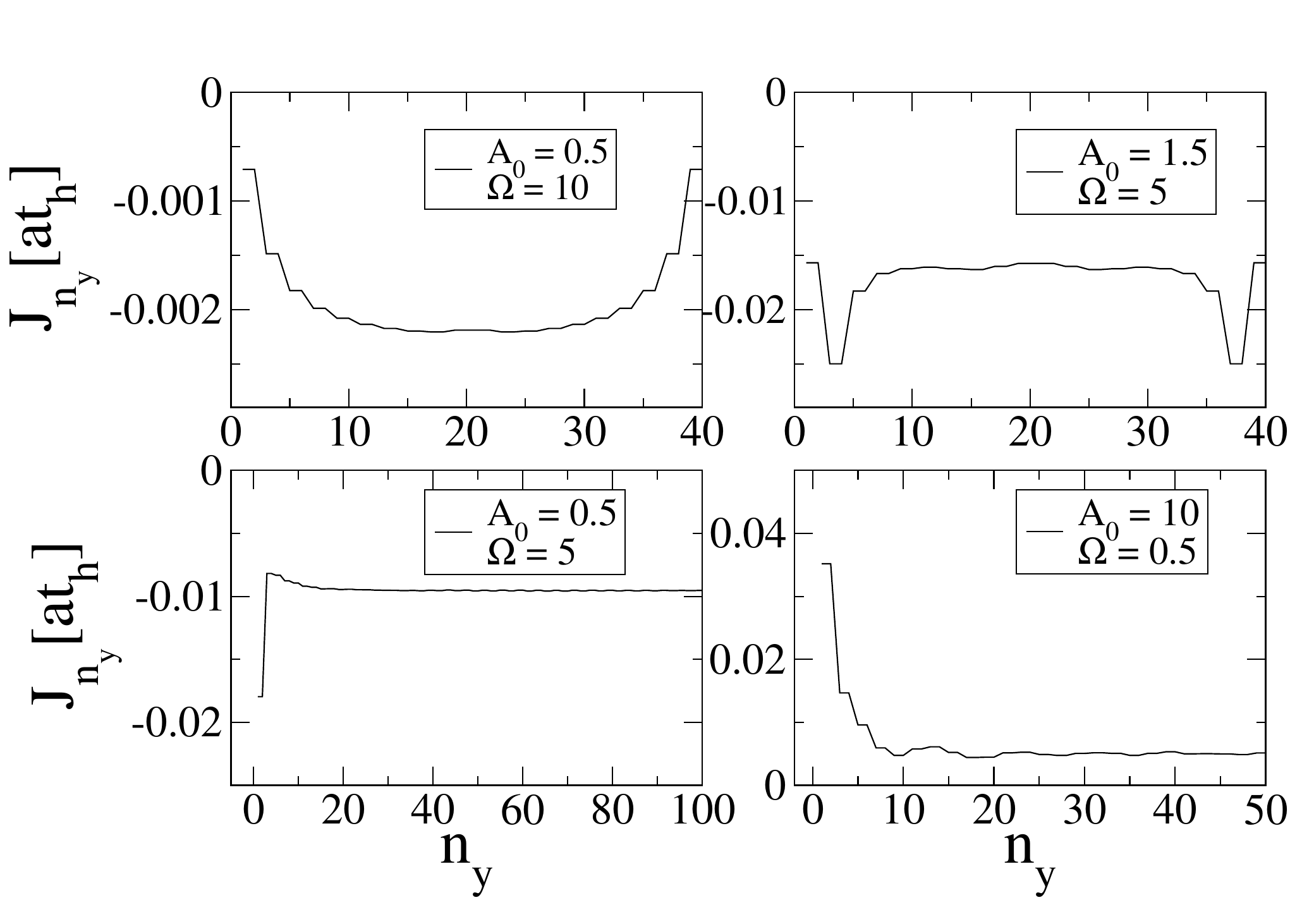}\\
\caption{
Current density flowing in the $\hat{x}$ direction and plotted as a function of the $y\equiv n_y$, 
at long times after the laser quench 
for clockwise from top-left 
$P_1\equiv \left[A_0a=0.5,\Omega = 10 t_h,C=1\right]$, 
$P_2\equiv \left[A_0a=1.5, \Omega = 5 t_h,C=1\right]$, $P_4\equiv \left[A_0a=10, \Omega=0.5 t_h,C=0\right]$, and
$P_3 \equiv \left[A_0a=0.5,\Omega=5t_h, C=3\right]$. The quench current is symmetric in $n_y$, hence
for the lower two panels, the current density is plotted over only half the strip.  
For a half-filled Floquet band, the current density is zero due to particle-hole symmetry.
}
\label{fig9}
\end{figure}

Figs.~\ref{fig7} and~\ref{fig8} show the current densities (all our results are time-averaged over
a laser cycle) for three Floquet eigenstates, one corresponding to the
lowest Floquet level $\alpha=1$, 
the second a level from
the middle of the lower band $\alpha=N_y/4$, and the third being the edge mode at $\alpha=N_y/2$. While Fig.~\ref{fig7} is 
for the two topological phases $P_{1,2}$ which correspond to a conventional Chern insulator,
Fig.~\ref{fig8} is for the cases $P_{3,4}$ 
where anomalous edge states appear at the 
Floquet zone boundaries. For Fig.~\ref{fig8} therefore, $\alpha=1$ is also an edge state. 
Note that since the edge state localization length for phases $P_{3,4}$ is quite long, for this
reason we had to work with cylinders of longer lengths so as to prevent the edge states at the boundaries of the 
FBZ from hybridizing.

As noted above, the current density of the exactly half-filled case is zero. The way this cancellation 
comes about for the Chern insulators $P_{1,2}$ (Fig.~\ref{fig7})
with a single edge state corresponding to $\alpha=N_y/2$ is that the current density of 
that edge state is opposite in sign to the current density of all the bulk states $\alpha = 1 \ldots (-1+N_y/2)$.
Thus while each bulk state contributes a relatively small amount to the current density, all their contributions add up to
a net value such that it exactly cancels the current density from the edge state.

For the phase $P_3$ the edge states from the FBZ boundaries ($\alpha=1$) have the opposite
chirality to the edge state from the FBZ center ($\alpha = N_y/2$), and this can be clearly seen in 
the top panel of Fig.~\ref{fig8}. This figure also shows that for $P_3$, the magnitude
of the edge-currents from the states in the zone-boundary are much smaller than those from the center.

For the phase $P_4$ the edge states from the FBZ boundaries ($\alpha=1$) have the same chirality
as the edge-state from the FBZ center ($\alpha = N_y/2$). This is reflected in the lower panel of 
Fig.~\ref{fig8} where
the peak values of the current densities from the edge states at $\alpha=1$ and $\alpha=N_y/2$ are 
indeed of the same sign, although they
may have opposite signs within a few lattice spacings of the boundary.

Note that even though current densities carried by the anomalous edge states at the FBZ boundaries can be of the same magnitude as
those of the edge states at the center of the FBZ, it does not imply that they affect physical observables in the same way.
This is because physical observables, such as the quench current density, are obtained by averaging over
all the Floquet states, where each state gets weighted by their respective occupation probabilities. 
Moreover note that this current density is not what one measures in transport such as Hall response, where the latter is a 
linear response to an external voltage difference, and given by the Kubo or Landauer formalism. 
As argued in the previous sub-section, 
an effective high temperature of the anomalous edge states imply they contribute relatively little to transport.

We now discuss the quench current density i.e., 
the current density carried by the wavefunction at long times after the quench. This is shown in 
the four panels in Fig.~\ref{fig9} for the four phases. 
Even though the current
density of each exact eigenstate is anti-symmetric in position along the cylinder, the quench current 
density is symmetric in position. 
\begin{eqnarray}
J_{N_y-n_y+1}(t\rightarrow \infty) = J_{n_y}(t\rightarrow\infty).
\end{eqnarray}
Why this is so is explained in Appendix~\ref{app3}. 
Here we give another quick way to understand this. 

It is convenient to define 
the time-averaged local density in the diagonal ensemble, which using Eq.~\eqref{nij} is,
\begin{eqnarray}
&&\rho_{n_y}= \frac{1}{N_x}\overline{\sum_{k_x} n_{n_y,n_y,k_x}(t)}\nonumber\\
&&= \frac{1}{N_x}\sum_{k_x,\alpha = 1 \ldots N_y}O_{\alpha}(k_x) \sum_m\biggl|\phi_{k_x,\alpha,n_y}^m\biggr|^2\label{rhodef}.
\end{eqnarray}
As shown in Appendix~\ref{app2}, the combination of being at half-filling, the particle-hole and inversion symmetry
of the Floquet Hamiltonian, and the fact that the quench breaks
inversion symmetry causing $O_{\alpha}(k_x) \neq O_{\alpha}(-k_x)$,   
results in a local deviation of the density from half-filling, which is anti-symmetric in position.
In particular, defining, $\delta \rho_{n_y}=\rho_{n_y}-1/2$, we have the relation,
\begin{eqnarray}
\delta \rho_{n_y}= -\delta \rho_{N_y-n_y+1}\label{cdev}.
\end{eqnarray}
Semi-classically, the current density is a product of $\delta \rho_{n_y} v_{n_y}$ 
where $v_{n_y}$ is the velocity. Since the perfect chirality of the exact Floquet eigenstates implies 
$v_{n_y}$ is anti-symmetric in position, and $\delta \rho_{n_y}$ is also anti-symmetric in position, 
this implies a net quench current density which is symmetric in position. 
If the quench had preserved inversion symmetry by providing an occupation probability with
the symmetry $O_{\alpha}(k_x)=O_{\alpha}(-k_x)$, then we would have had $\delta\rho_{n_y}=0$, and consequently the
time-averaged quench current density would have been zero at half-filling.

A simple way to see why all this comes about due to the breaking of inversion symmetry, note that at the time
of the laser quench at $t=0$, the vector potential is completely pointing along the $\hat{x}$ direction. Thus
inversion symmetry in $\hat{x}$ is broken resulting in an unequal population of states at $+k_x$ and $-k_x$.
This implies that for every $|k_x|$, there is a net current flowing in the system due to periodic boundary
conditions in $\hat{x}$. Since individual eigenstates are exactly chiral along the $\hat{y}$ direction,
this current density has to also imply that some net charge be moved from one end of the cylinder to the other.
A slower quench will reduce the magnitude of this effect.   

To generate this quench current density, all we needed was to break inversion symmetry. 
If we had no particle-hole symmetry for example due to next-nearest-neighbor hopping, then we would have still
generated a quench current density, but this current density would not have been exactly symmetric in position.

Thus we find that the quench current density appears as a sheet of circulating current on the surface of the cylinder. 
Such a current density profile will generate a
local magnetization, and is therefore detectable using sensitive magnetometers such as SQUIDs~\cite{Bluhm09,Shibata15}.
This result may also be useful for using
a fast quench as a tool to generate dissipationless current flow in a carbon nanotube.

We now briefly discuss how our results depend on the the length of the cylinder. We have taken care to ensure that the
length is sufficiently long so as to clearly identify the bulk and edge states. As the length is further increased
from our chosen lengths, the edge states are not modified, while the bulk spectrum fills out more
as more states are being added. The physical observables we study are properly normalized to account for this effect, and therefore 
do not depend sensitively on the length of the cylinder. 

\section{Conclusions}\label{conclu}

One of the properties that make FTIs unique is a rich structure of edge states, with edge modes
appearing both at the center of the FBZ as well as the boundaries. For
a low amplitude laser, one can identify the former 
with off-resonant, and the latter by resonant processes. In this paper
we have highlighted how these qualitatively different edge modes behave, the current densities
carried by them, and how they are occupied 
in a closed quantum system where the laser was switched on as a quench. 

We find that for an off-resonant laser,
despite the laser quench, the Floquet level occupation is remarkably close to a half-filled zero temperature Fermi
function. This is consistent with a bulk Kubo-formula computation for the Hall conductance~\cite{Dehghani15a} which
found it to be quite close to $e^2/h$. 
For the resonant laser, we find a selective depletion and occupation
of the Floquet modes, i.e., a laser induced population inversion for
selected regions in momentum space. As a consequence we find that in the same phase (phase $P_3$ for example)
the edge states at the center of the FBZ are occupied at a low effective temperature, while the
edge states at the boundaries of the FBZ, and arising due to resonant processes, are occupied with
a high effective temperature. 
A slower quench will not qualitatively affect this result
because the laser resonance condition does not depend on the amplitude of the laser, but only on its frequency~\cite{Santoro15}. 

By simply looking at how the edge states are occupied, and what fraction
of the bulk is excited, we can use the Landauer formalism to make a simple estimate for the conductance of the edge states and hence 
the Hall response. We find this
estimate to be consistent with the 
Hall response for a spatially uniform system where
no edge modes exist and the entire Hall response is purely due to bulk states
and was computed using the Kubo formalism~\cite{Dehghani15a,Dehghani15b}. 
This is a signature of the bulk-boundary correspondence in FTIs where Hall response can be 
captured by two complementary ways, one entirely involving bulk states in an infinite system,
and the second involving effective 1D transport along chiral edge states. Moreover this correspondence
and in particular the edge state picture 
also explains why the Hall response for some phases like $P_3$, when accounting properly for the
nonequilibrium occupation is only $\sim 1/3$ of its maximum possible value of $C e^2/h$.

We also find that the  
expectation values of the time-averaged quench current density shows some special symmetries, for
example it is exactly symmetric along the length of the cylinder. We have shown that these
have to do with the underlying
particle-hole and inversion symmetry of the Floquet Hamiltonian and of graphene, where for the former the inversion
symmetry manifests itself only 
after time-averaging over one cycle of the laser. 

We also showed that the quench current density arises because the laser quench breaks inversion symmetry say in the $x$-direction, leading to
an asymmetric occupation of $+k_x$ and $-k_x$ states. Thus when periodic boundary conditions are imposed
in the $x$-direction, this leads to a circulating current on the surface of the cylinder. This
current density profile is markedly different from the naive expectation of having clockwise currents at the
top, and anti-clockwise currents at the bottom of the cylinder, where the latter would be the profile
only in an exact eigenstate of the system. Since each eigenstate is chiral, we also showed that the quench current density 
profile leads to a removal of charge from one end of the cylinder to the other. This unusual current density profile
can be detected using magnetometers such as SQUIDs. Moreover laser quenches can be used as a tool
for generating a net dissipationless current flow in carbon nanotubes.

{\sl Acknowledgments:}
The authors thank Y. Lemonik and M. Rudner for helpful discussions.
This work was supported by the US Department of Energy,
Office of Science, Basic Energy Sciences, under Award No.~DE-SC0010821.

\appendix

\section{
Particle-hole symmetry of occupation probabilities 
(proof of Eq.~\eqref{DistFuncCompl})}
\label{app1}
The proof of this equation is based on the particle-hole symmetry of graphene and the Floquet Hamiltonians. 
According to this symmetry, energy eigenvalues appear in pairs of opposite signs, and the
corresponding eigenstates are not independent, but can be obtained from one another. 
This relation in general depends on the representation of the Hamiltonian and in our geometry it has the
following common 
form for graphene ($|a\rangle$) and Floquet eigenstates ($|\phi\rangle $), 
\begin{subequations}
\begin{align}
\phi_{k_x,\alpha,n_y}(t)&=(-1)^{n_y}\phi_{k_x,N_y - \alpha + 1,N_y - n_y + 1}^*(t),\\
a_{k_x,l,n_y}&=(-1)^{n_y}a_{k_x,N_y - l + 1,N_y - n_y + 1}^*.
\end{align}\label{phrel}
\end{subequations}
Among the three subscripts, the second label $\alpha$ denotes energy/quasi-energy while the third label $n_y$ denotes position along 
the cylinder. There are a total of $N_y$ values for each of these labels. 
We now justify Eq.~\eqref{phrel}, and then use it to derive Eq.~\eqref{DistFuncCompl}.

Both graphene and the time-dependent Hamiltonian $H(t)$ have only nearest-neighbor hopping.
Thus an operation ${\cal P}$ which simply changes the sign of the wave function on the $B$ sublattice relative to
the $A$ sublattice, is equivalent to reversing the sign of the Hamiltonian (in this
appendix the Hamiltonian is the first quantized version of Eq.~\eqref{H1} for graphene and Eq.~\eqref{Hedge1} for the time-periodic system), 
\begin{eqnarray}
{\cal P} H(k_x,t) {\cal P}^{-1} = -H(k_x,t).
\end{eqnarray}
Let us now consider a second operation ${\cal I}_y$ 
that inverts the system along the length of the cylinder, {\sl i.e.},  interchanges sites $n_y\Leftrightarrow N_y-n_y+1$. 
Under this operation, 
\begin{eqnarray}
{\cal I}_y H(k_x,t) {\cal I}^{-1}_y = H^*(k_x,t)\label{IyH}.
\end{eqnarray}
The above two transformation can be combined into a single one which is simply the charge-conjugation
or particle-hole transformation
\begin{eqnarray}
\biggl({\cal P I}_y\biggr) H^*(k_x,t) \biggl({\cal P I}_y\biggr)^{-1} = -H(k_x,t)\label{phsym}.
\end{eqnarray}
Now the Floquet modes obey the Schr\"odinger equation for the Floquet Hamiltonian $H-i\partial_t$
\begin{eqnarray}
\biggl[H(t) -i\partial_t -\epsilon\biggr]\phi(t) = 0\label{sym1}.
\end{eqnarray}
Let us complex-conjugate the above equation,
\begin{eqnarray}
\biggl[H^*(t) +i\partial_t -\epsilon\biggr]\phi^*(t) = 0.
\end{eqnarray}
Now let us apply the operation ${\cal I}_y$,
\begin{eqnarray}
&&\biggl[{\cal I}_y H^*(t){\cal I}^{-1}_y +i\partial_t -\epsilon\biggr]{\cal I}_y\phi^*(t) \nonumber\\
&&=\biggl[H(t) +i\partial_t -\epsilon\biggr]{\cal I}_y\phi^*(t) = 0.
\end{eqnarray}
Next we apply ${\cal P}$ to the above to obtain,
\begin{eqnarray}
&&\biggl[{\cal P}H(t){\cal P}^{-1} +i\partial_t -\epsilon\biggr]{\cal P I}_y\phi^*(t)=0, \nonumber\\
&&\biggl[-H(t) +i\partial_t -\epsilon\biggr]{\cal P I}_y\phi^*(t) = 0.
\end{eqnarray}
The above implies
\begin{eqnarray}
&&\biggl[H(t) -i\partial_t +\epsilon\biggr]{\cal P I}_y\phi^*(t) = 0\label{sym2}.
\end{eqnarray}
Thus Eqs.~\eqref{sym1},~\eqref{sym2} together give that if $\phi(t)$ is a Floquet mode with 
quasi-energy $\epsilon$, then ${\cal P I}_y\phi^*(t)$ is a Floquet mode with the opposite quasi-energy 
$-\epsilon$. Thus the components of $\phi$ in a cylindrical geometry, which we denote by $\phi_{k_x,\alpha,n_y}$,
with $k_x$ labeling momentum, $\alpha$ labeling quasi-energy, and $n_y$ labeling position along the cylinder,
must be related as Eq.~\eqref{phrel}

In the following we drop $k_x$ in the subscript to shorten the equations.
Since in the definition of $O_{\alpha}(k_x)$ there is a summation over the occupied energies of graphene, 
we also need the completeness equation of these eigenstates 
\begin{eqnarray}
\sum_{l=1 \cdots N_y}|a_{l}\rangle \langle a_{l}| = {\bf 1},
\end{eqnarray}
where ${\bf 1}$ is the $N_y \times N_y$ identity matrix. 
We can split this sum into the upper half $(l = N_y/2 +1 \ldots N_y)$ which includes
unoccupied states and the lower half $(l = 1 \ldots N_y/2)$ which includes occupied states
\begin{eqnarray}
\sum_{l={\rm unocc}}a_{l, n_y}a_{l, n_y'}^*+\sum_{l={\rm occ}} a_{l, n_y}a_{l, n_y'}^* = \delta_{n_y,n_y'}.
\end{eqnarray}
Applying the particle-hole symmetry to the first sum we obtain
\begin{eqnarray}
\sum_{l={\rm unocc}}(-1)^{n_y+n_y'}a_{N_y-l+1, N_y-n_y+1}^*&&a_{N_y-l+1, N_y-n_y'+1}\notag\\
+\sum_{l={\rm occ}} a_{l, n_y}a_{l, n_y'}^*&&= \delta_{n_y,n_y'}.
\end{eqnarray}
Since $l$ is a dummy index in the sum, we relabel $l$ in the first sum with $N_y-l+1$. 
Therefore both sums will be over occupied states and we obtain, 
\begin{eqnarray}
\sum_{l={\rm occ}}&&a_{l, n_y}a_{l, n_y'}^* = \delta_{n_y,n_y'} \notag\\
&&\!\!-\sum_{l={\rm occ}}(-1)^{n_y+n_y'}a_{l, N_y-n_y+1}^*a_{l, N_y-n_y'+1}\label{halfSum}.
\end{eqnarray}
Now we can apply this result to prove the required relation for $O_{\alpha}(k_x)$. 
Starting with the definition of $O_{N_y-\alpha+1}(k_x)$ we have 
\begin{eqnarray}
O_{N_y-\alpha+1}(k_x) = &&\!\!\sum_{n_y,n_y'}\!\!\ \phi_{N_y-\alpha+1, n_y}^*(0)
\phi_{N_y-\alpha+1, n_y'}(0)\notag\\ &&\times\sum_{l={\rm occ}}a_{l,n_y}a^*_{l,n_y'}.
\end{eqnarray}
Now we apply the particle-hole symmetry on the Floquet states to get
\begin{eqnarray}
&&O_{N_y-\alpha+1}(k_x) = \nonumber\\
&&\sum_{n_y,n_y'}\!\!\ (-1)^{n_y+n_y'}\phi_{\alpha, N_y-n_y+1}(0)\phi_{\alpha, N_y-n_y'+1}^*(0)\notag\\ &&\times\sum_{l={\rm occ}}a_{l,n_y}a^*_{l,n_y'}.
\end{eqnarray}
We can insert Eq.~\eqref{halfSum} in the above to obtain
\begin{eqnarray}
&&O_{N_y-\alpha+1}(k_x) = \!\!\sum_{n_y,n_y'}\!\!\ \phi_{\alpha, N_y-n_y+1}(0)
\phi_{\alpha, N_y-n_y'+1}^*(0)\notag\times\\&&\biggl[(-1)^{n_y+n_y'}\delta_{n_y,n_y'} -\sum_{l={\rm occ}}a_{l, N_y-n_y+1}^*a_{l, N_y-n_y'+1}\biggr].
\end{eqnarray}
After relabeling the sum variables $n_y$ and $n_y'$, 
and performing some straightforward algebra, one finds that this equation can be rewritten in the following bra-ket notation 
\begin{eqnarray}
&&O_{N_y-\alpha+1}(k_x) = \langle\phi_{k_x,\alpha}(0)|\phi_{k_x,\alpha}(0)\rangle\notag\\
&& -\sum_{l={\rm occ}}\langle\phi_{k_x,\alpha}(0)|a_{k_x,l}\rangle \langle a_{k_x,l}|\phi_{k_x,\alpha}(0)\rangle,
\end{eqnarray}
where we have restored the $k_x$ indices again. This proves the desired equation
\begin{eqnarray}
&&O_{N_y-\alpha+1}(k_x) = 1 -O_{\alpha}(k_x).
\end{eqnarray}

\section{Anti-symmetry of density deviations (proof of Eq.~\eqref{cdev})}
\label{app2}
To prove Eq.~\eqref{cdev}, in addition to Eq.~\eqref{phsym} which implies a particle hole symmetry of the Hamiltonian spectrum, 
we need another relation which relates the eigenstates of the Hamiltonian at $k_x$ and $-k_x$. 
This can be done using the properties of the system under inversion  around the center of one of the hexagons.
Under inversion in both $x$ and $y$ directions, 
\begin{eqnarray}
H(k_x, t) = H^*(-k_x, t + \pi/\Omega).\label{HInv}
\end{eqnarray}
Note that since the momentum is odd under inversion symmetry, on the right the wave vector argument is $-k_x$. 
This also explains why the time argument of the Hamitonian on the right, is shifted by $\pi/\Omega$. 
This is because our circularly polarized field obeys $\vec{A}(t+\pi/\Omega) = -\vec{A}(t)$ which is what we need for inversion 
symmetry to be satisfied. Therefore, although inversion symmetry is not satisfied between Hamiltonians at equal times, 
a weaker version of this symmetry which relates two Hamiltonians at different times, still holds. 

Now consider the Floquet equation after the operation ${\cal I}_{y}$
\begin{eqnarray}
\biggl[{\cal I}_{y}H(k_x, t){\cal I}_{y}^{-1} -i\partial_t -\epsilon\biggr]{\cal I}_{y}\phi(k_x, t) = 0\label{FloqEqIy}.
\end{eqnarray}
Inserting Eq.~\eqref{IyH} in the above, one finds
\begin{eqnarray}
\biggl[H^*(k_x, t) -i\partial_t -\epsilon\biggr]{\cal I}_{y}\phi(k_x, t) = 0.
\end{eqnarray}
Now we use Eq.~\eqref{HInv} to obtain
\begin{eqnarray}
\biggl[H(-k_x, t+\pi/\Omega) -i\partial_t -\epsilon\biggr]{\cal I}_{y}\phi(k_x, t) = 0.
\end{eqnarray}
That is to say we can identify ${\cal I}_{y}\phi(k_x, t)$ with $\phi(-k_x, t+\pi/\Omega)$ up to a phase factor. In components this becomes
\begin{eqnarray}
\phi_{k_x,\alpha, n_y}(t)=\phi_{-k_x,\alpha, N_y - n_y + 1}(t+\pi/\Omega).\label{PhiInv}
\end{eqnarray}
After squaring this equation, we get
\begin{eqnarray}
|\phi_{k_x,\alpha, n_y}(t)|^2=|\phi_{-k_x,\alpha, N_y - n_y + 1}(t+\pi/\Omega)|^2.
\end{eqnarray}
Now as usual we can expand the Floquet eigenstates in 
their Fourier harmonics, $|\phi_{\alpha}(t)\rangle=\sum_m|\phi_{\alpha}^m\rangle e^{im\Omega t}$.  This will give
\begin{eqnarray}
&&\sum_{m,m'}\phi_{k_x,\alpha, n_y}^{m}\phi_{k_x,\alpha, n_y}^{m-m'*}e^{im'\Omega t}=\notag\\&&
\sum_{m,m'}(-1)^{m'} \phi_{-k_x,\alpha, N_y - n_y + 1}^{m}\phi_{-k_x,\alpha, N_y - n_y + 1}^{m-m'*}e^{i m'\Omega t}.
\end{eqnarray}
Since the Fourier harmonics with different exponents are orthogonal, it is possible to equate the coefficients of the 
same harmonics on the two sides. However, for our purpose, only $m'=0$ would be required. Therefore we define
\begin{eqnarray}
S_{\alpha,n_y}(k_x)=\sum_m \biggl|\phi^m_{k_x,\alpha,n_y}\biggr|^2,
\end{eqnarray}
where according to the above reasoning, one finds
\begin{eqnarray}
S_{\alpha,n_y}(k_x)= S_{\alpha,N_y-n_y +1}(-k_x).\label{inq}
\end{eqnarray}
The Floquet Hamiltonian also has particle-hole symmetry which relates 
the energy eigenvalues and eigenfunctions of the lower and upper Floquet bands at the same
$k_x$ and at every instant of time. As noted in App.~\ref{app1} this symmetry imposes
\begin{eqnarray}
\phi_{k_x,\alpha,n_y}(t)=(-1)^{n_y}\phi_{k_x,N_y - \alpha + 1,N_y - n_y + 1}^*(t).
\end{eqnarray}
We can square this relation to obtain
\begin{eqnarray}
|\phi_{k_x,\alpha,n_y}(t)|^2=|\phi_{k_x,N_y - \alpha + 1,N_y - n_y + 1}(t)|^2.
\end{eqnarray}
As for the inversion symmetry, here too 
we can expand the two sides of this equation in Fourier harmonics and after equating the coefficients obtain
\begin{eqnarray}
S_{\alpha,n_y}(k_x)= S_{N_y-\alpha +1,N_y-n_y+1}(k_x)\label{phq}.
\end{eqnarray}

We may write the local density after the quench in the diagonal ensemble, defined in Eq.~\eqref{rhodef}, as
\begin{eqnarray}
\rho_{n_y}=\frac{1}{N_x}\sum_{k_x,\alpha}O_{\alpha}(k_x)S_{\alpha,n_y}(k_x)\label{apq1},
\end{eqnarray}
where $N_x$ denotes the total number of points in the $k_x$ sum.
As shown in App.~\ref{app1}, particle-hole symmetry both for graphene and the Floquet eigenstates give,
\begin{eqnarray}
O_{\alpha}(k_x) = 1 -O_{N_y-\alpha + 1}(k_x).
\end{eqnarray}
Changing variables in Eq.~\eqref{apq1} from $\alpha \rightarrow N_y -\alpha +1$, and using the above relation for
$O_{\alpha}$, we  get,
\begin{eqnarray}
&&\rho_{n_y}=\frac{1}{N_x}\sum_{k_x,\alpha}O_{N_y-\alpha+1}(k_x)S_{N_y-\alpha+1,n_y}(k_x)\nonumber\\
&&=\frac{1}{N_x}\sum_{k_x,\alpha}\biggl[1-O_{\alpha}(k_x)\biggr]S_{N_y-\alpha+1,n_y}(k_x).
\end{eqnarray}
Using Eq.~\eqref{phq}
\begin{eqnarray}
&&\rho_{n_y}=\frac{1}{N_x}\sum_{k_x,\alpha}\biggl[1-O_{\alpha}(k_x)\biggr]S_{\alpha,N_y-n_y+1}(k_x)\nonumber\\
&&= \frac{1}{N_x}\sum_{k_x,\alpha}S_{\alpha,N_y-n_y+1}(k_x) -\rho_{N_y-n_y+1}\label{rhq2}.
\end{eqnarray}

At every instant of time, the Floquet eigenstates form a complete basis. Thus
\begin{eqnarray}
\sum_{\alpha}|\phi_{k_x,\alpha}(t)\rangle \langle \phi_{k_x,\alpha}(t)| = {\bf 1},
\end{eqnarray}
where ${\bf 1}$ is the $N_y \times N_y$ identity matrix. After a Fourier expansion we obtain
\begin{eqnarray}
\sum_{\alpha,m,m'}e^{i(m-m')\Omega t}\phi_{k_x,\alpha,n_y}^m \biggl[\phi_{k_x,\alpha,n_y'}^{m'}\biggr]^*= \delta_{n_y,n_y'}.
\end{eqnarray}
Since the r.h.s.~is time-independent, the only non-vanishing terms arise for $m=m'$. 
Thus,
\begin{eqnarray}
\sum_{\alpha,m}\phi_{k_x,\alpha,n_y}^m \biggl[\phi_{k_x,\alpha,n_y'}^{m}\biggr]^*= \delta_{n_y,n_y'},
\end{eqnarray}
or,
\begin{eqnarray}
\sum_{\alpha,m}\phi_{k_x,\alpha,n_y}^m \biggl[\phi_{k_x,\alpha,n_y}^{m}\biggr]^*= 1.
\end{eqnarray}
The above implies,
\begin{eqnarray}
\sum_{\alpha}S_{\alpha,n_y}(k_x) = 1,
\end{eqnarray}
and similarly for the momentum average of the above quantity, we may write,
\begin{eqnarray}
\frac{1}{N_x}\sum_{k_x,\alpha}S_{\alpha,n_y}(k_x) = 1.
\end{eqnarray}
Substituting this in Eq.~\eqref{rhq2}, we obtain,
\begin{eqnarray}
\rho_{n_y}+\rho_{N_y-n_y+1}=1\label{ap3}.
\end{eqnarray}
Writing $\rho_{n_y}= 1/2 +\delta \rho_{n_y}$, this immediately implies,
\begin{eqnarray}
\delta \rho_{n_y}= - \delta \rho_{N_y-n_y+1}.
\end{eqnarray}
Thus we have proved Eq.~\eqref{cdev}.

Now we show that if the quench had preserved inversion symmetry $O_{\alpha}(k_x)=O_{\alpha}(-k_x)$,
then there would have been no local deviation from half-filling  i.e., $\delta \rho_{n_y}=0$. 
To prove this, we start by writing
\begin{eqnarray}
\rho_{n_y}=\frac{1}{N_x}\sum_{k_x,\alpha}O_{\alpha}(-k_x)S_{\alpha,n_y}(-k_x)\label{apq2}.
\end{eqnarray} 
Using the inversion symmetry of the Floquet eigenmodes in Eq.~\eqref{inq}, the above implies,
\begin{eqnarray}
\rho_{n_y}=\frac{1}{N_x}\sum_{k_x,\alpha}O_{\alpha}(-k_x)S_{\alpha,N_y-n_y+1}(k_x).
\end{eqnarray}
If $O_{\alpha}(k_x)=O_{\alpha}(-k_x)$, the above leads to
$\rho_{n_y} = \rho_{N_y-n_y+1}$.
But we also have the constraint in Eq.~\eqref{ap3}, so that for this case $\delta \rho_{n_y}=0$. 
Thus if $O_{\alpha}(k_x)=O_{\alpha}(-k_x)$, this would lead to 
no local deviation from half-filling. As we show in the next appendix, this will also lead
to a zero time-averaged quench current density
at half-filling.

\section{Symmetries of the current density of Floquet eigenstates and the quench
current density}\label{app3}

In this section, using the symmetry relations of the eigenstates obtained in previous appendices, we prove Eqs.~\eqref{CurrentInv} and
~\eqref{CurrentChiral}.  We also prove that the quench current density is symmetric i.e., $J_{n_y}=J_{N_y-n_y+1}$,
where $J_{n_y}$ is defined in Eq.~\eqref{currq}.

First let us consider Eq.~\eqref{CurrentInv}. To derive this equation, we will apply the inversion symmetry properties 
of the eigenstates i.e., Eq.~\eqref{PhiInv}.
By Fourier expanding the two sides of this equation, one finds
\begin{eqnarray}
\phi_{k_x,\alpha, n_y}^{m}=(-1)^{m}\phi_{-k_x,\alpha, N_y - n_y + 1}^{m}.\label{PhiInv2}
\end{eqnarray}
Consider the current density carried by a Floquet eigenstate $\alpha$ (Eq.~\eqref{curres},~\eqref{curres2a}). 
Under inversion in $j_{\alpha, 2n_y - 1}(k_x)$, its arguments $k_x$ and $2n_y - 1$, would be transformed to $-k_x$ and $N_y - 2n_y + 2 $. 
Therefore we consider the following,
\begin{eqnarray}
&&j_{\alpha,N_y - 2n_y + 2}(-k_x)=\nonumber\\
&&\sqrt{3}t_h a\sum_{m,n}\biggl[\tilde{J}_{-m}\left(A_0a\right)\sin\biggl(\frac{-\sqrt{3}k_x a}{2}-\frac{m\pi}{3}\biggr)\biggr]
\nonumber\\
&&\times 2{\rm Re}\biggl[\phi_{-k_x,\alpha,N_y - 2n_y + 2}^n\Bigl(\phi_{-k_x,\alpha,N_y - 2n_y + 1}^{n+m}\Bigr)^* \biggr].
\end{eqnarray}
We insert Eq.~\eqref{PhiInv2} above to obtain,
\begin{eqnarray}
&&j_{\alpha,N_y - 2n_y + 2}(-k_x)=\nonumber\\
&&\sqrt{3}t_h a\sum_{m,n}(-1)^{m}\biggl[\tilde{J}_{-m}\left(A_0a\right)\sin\biggl(\frac{-\sqrt{3}k_x a}{2}-\frac{m\pi}{3}\biggr)\biggr]
\nonumber\\
&&\times 2{\rm Re}\biggl[\phi_{k_x,\alpha,2n_y-1}^{n}\Bigl(\phi_{k_x,\alpha,2n_y}^{n+m}\Bigr)^{*} \biggr].
\end{eqnarray}
After using $\tilde{J}_{-m} = (-1)^{m}\tilde{J}_{m}$, one can relabel $m$ and $n$, to obtain
\begin{eqnarray}
&&j_{\alpha,N_y - 2n_y + 2}(-k_x)=\nonumber\\
&&\sqrt{3}t_h a\sum_{m,n}\biggl[\tilde{J}_{-m}\left(A_0a\right)\sin\biggl(\frac{-\sqrt{3}k_x a}{2}+\frac{m\pi}{3}\biggr)\biggr]
\nonumber\\
&&\times 2{\rm Re}\biggl[\phi_{k_x,\alpha,2n_y}^{n}\Bigl(\phi_{k_x,\alpha,2n_y-1}^{n+m}\Bigr)^{*} \biggr].
\end{eqnarray}
Comparison of the above result and the definition of $j_{\alpha,n_y}(k_x)$, we find 
\begin{eqnarray}
j_{\alpha,n_y}(-k_x) = -j_{\alpha,N_y-n_y+1}(k_x),\label{antij1}
\end{eqnarray}
which after summing over $k_x$ proves that the average current density 
carried by a Floquet eigenstate is anti-symmetric in position $J_{\alpha,n_y}=-J_{\alpha, N_y-n_y+1}$. 

Next, we prove Eq.~\eqref{CurrentChiral}. 
For this we  need to Fourier transform Eq.~\eqref{phrel}
\begin{eqnarray}
\phi_{k_x,\alpha,n_y}^{m}&=(-1)^{n_y}\Bigl(\phi_{k_x,N_y - \alpha + 1,N_y - n_y + 1}^{-m}\Bigr)^*.\label{phiPHFourier}
\end{eqnarray}
Combination of the above with Eq.~\eqref{PhiInv2} gives
\begin{eqnarray}
\phi_{k_x,\alpha,n_y}^{m}&=(-1)^{m+n_y}\Bigl(\phi_{-k_x,N_y - \alpha + 1,n_y}^{-m}\Bigr)^*.\label{phrelFourier}
\end{eqnarray}
According to the definition of the current-density in a Floquet eigenstate, we have
\begin{eqnarray}
&&j_{N_y - \alpha + 1,2n_y-1}(k_x)=\nonumber\\
&&\sqrt{3}t_h a\sum_{m,n}\biggl[\tilde{J}_{-m}\left(A_0a\right)\sin\biggl(\frac{\sqrt{3}k_x a}{2}-\frac{m\pi}{3}\biggr)\biggr]
\nonumber\\
&&\times 2{\rm Re}\biggl[\phi_{k_x,N_y - \alpha + 1,2n_y}^n\Bigl(\phi_{k_x,N_y - \alpha + 1,2n_y-1}^{n+m}\Bigr)^* \biggr].
\end{eqnarray}
Insertion of Eq.~\eqref{phrelFourier} in the above gives 
\begin{eqnarray}
&&j_{N_y - \alpha + 1,2n_y-1}(k_x)=\nonumber\\
&&-\sqrt{3}t_h a\sum_{m,n}(-1)^{m}\biggl[\tilde{J}_{-m}\left(A_0a\right)\sin\biggl(\frac{\sqrt{3}k_x a}{2}-\frac{m\pi}{3}\biggr)\biggr]
\nonumber\\
&&\times 2{\rm Re}\biggl[\Bigl(\phi_{-k_x,\alpha,2n_y}^{-n}\Bigr)^*\phi_{-k_x,\alpha,2n_y-1}^{-n-m} \biggr].
\end{eqnarray}
Applying $\tilde{J}_{-m} = (-1)^{m}\tilde{J}_{m}$ and relabeling $m$ and $n$ gives 
\begin{eqnarray}
&&j_{N_y - \alpha + 1,2n_y-1}(k_x)=\nonumber\\
&&-\sqrt{3}t_h a\sum_{m,n}\biggl[\tilde{J}_{-m}\left(A_0a\right)\sin\biggl(\frac{\sqrt{3}k_x a}{2}+\frac{m\pi}{3}\biggr)\biggr]
\nonumber\\
&&\times2{\rm Re}\biggl[\Bigl(\phi_{-k_x,\alpha,2n_y}^{n}\Bigr)^*\phi_{-k_x,\alpha,2n_y-1}^{n+m} \biggr].
\end{eqnarray}
After absorbing the minus sign, we find
\begin{eqnarray}
&&j_{N_y - \alpha + 1,2n_y-1}(k_x)=\nonumber\\
&&\sqrt{3}t_h a\sum_{m,n}\biggl[\tilde{J}_{-m}\left(A_0a\right)\sin\biggl(\frac{-\sqrt{3}k_x a}{2}-\frac{m\pi}{3}\biggr)\biggr]
\nonumber\\
&&\times2{\rm Re}\biggl[\Bigl(\phi_{-k_x,\alpha,2n_y}^{n}\Bigr)^*\phi_{-k_x,\alpha,2n_y-1}^{n+m} \biggr].
\end{eqnarray}
From above it is evident that
\begin{eqnarray}
&&j_{N_y - \alpha + 1,n_y}(k_x)=j_{\alpha,n_y}(-k_x).\label{symj2}
\end{eqnarray}
After summing over $k_x$, we prove that pairs of Floquet eigenstates 
with quasi-energy $\epsilon,-\epsilon$ carry the same average current density,
\begin{eqnarray}
&&J_{N_y - \alpha + 1,n_y}=J_{\alpha,n_y}.
\end{eqnarray}

Next we prove that the quench current in Eq.~\eqref{currq} is symmetric in position. 
For this we start with the definition,
\begin{eqnarray}
J_{n_y}(t\rightarrow \infty)= \frac{1}{N_x}\sum_{k_x,\alpha={1\ldots N_y} }O_{\alpha}(k_x)j_{\alpha,n_y}(k_x)\label{currq1},
\end{eqnarray}
and substitute Eq.~\eqref{DistFuncCompl}, above. Then using that the current in a fully occupied band vanishes,
we obtain,
\begin{eqnarray}
J_{n_y} = -\frac{1}{N_x}\sum_{k_x,\alpha}O_{N_y-\alpha +1}(k_x)j_{\alpha,n_y}(k_x).
\end{eqnarray}
Now we change variables $n_y \rightarrow N_y-n_y+1$ to obtain,
\begin{eqnarray}
\!\!J_{N_y-n_y+1} \!\!= -\frac{1}{N_x}\sum_{k_x,\alpha}O_{N_y-\alpha +1}(k_x)j_{\alpha,N_y-n_y+1}(k_x).\nonumber\\
\label{sym3}
\end{eqnarray}
Next we use the recently proved relations in Eq.~\eqref{antij1} and~\ref{symj2} to show that
\begin{eqnarray}
j_{\alpha,N_y-n_y+1}(k_x) &&= - j_{\alpha,n_y}(-k_x)\nonumber\\
&&=-j_{N_y-\alpha+1,n_y}(k_x).
\end{eqnarray}
Substituting the above in Eq.~\eqref{sym3} and relabeling the labels $N_y-\alpha+1 \rightarrow \alpha$,
we prove that the quench current density is symmetric in position,
\begin{eqnarray}
J_{N_y-n_y+1} &&= \frac{1}{N_x}\sum_{k_x,\alpha}O_{\alpha }(k_x)j_{\alpha,n_y}(k_x)\nonumber\\
&&= J_{n_y}.
\end{eqnarray}

Now we show that if $O_{\alpha}(k_x)$ was symmetric in momentum space, $O_{\alpha}(k_x)= O_{\alpha}(-k_x)$,
the quench current density would vanish.
From using Eq.~\eqref{antij1}, and using the definition of the quench current density, we may write, 
\begin{eqnarray}
J_{n_y}&&= \frac{1}{N_x}\sum_{\alpha, k_x}O_{\alpha}(k_x) j_{\alpha,n_y}(k_x)\nonumber\\
&&= - \frac{1}{N_x}\sum_{\alpha, k_x}O_{\alpha}(k_x) j_{\alpha,N_y-n_y+1}(-k_x).
\end{eqnarray}
But an occupation which is symmetric in $k_x$ will allow us to rewrite the above as,
\begin{eqnarray}
J_{n_y}&&= -\frac{1}{N_x}\sum_{\alpha, k_x}O_{\alpha}(k_x) j_{\alpha,N_y-n_y+1}(k_x)\nonumber\\
&&=-J_{N_y-n_y+1},
\end{eqnarray}
implying a quench current density which is anti-symmetric in position. Since the
quench current density cannot be both symmetric and anti-symmetric in position, it implies
that if $O_{\alpha}(k_x)=O_{\alpha}(-k_x)$, this would lead to a zero quench current density.
Thus the breaking of the inversion symmetry by the switch-on of the laser is required to generate
a net quench current at half filling.

%

\end{document}